\documentclass[11pt, a4paper]{article}

\usepackage{amsmath}
\usepackage{amssymb}
\usepackage{amsthm}
\usepackage{amscd}
\usepackage{stmaryrd}
\usepackage{txfonts} 
\usepackage{tikz}
\usetikzlibrary{calc}

\usepackage[auth-sc, affil-it]{authblk}

\usepackage[utf8]{inputenc}
\usepackage[english]{babel}
\usepackage[T1]{fontenc}
\usepackage{lmodern}
\usepackage{microtype}
\usepackage[
    linktocpage=true,
    colorlinks=true,
    linkcolor=purple,
    citecolor=black,
    urlcolor=purple,
    bookmarksnumbered=true,
]{hyperref}

\usepackage[totalwidth=160mm, totalheight=247mm]{geometry}

\theoremstyle{plain}
\newtheorem{theo}{Theorem}[section]
\newtheorem{prop}[theo]{Proposition}

\theoremstyle{definition}

\newtheorem{exam}[theo]{Example}
\newtheorem{rema}[theo]{Remark}

\renewcommand{\leq}{\leqslant}
\renewcommand{\geq}{\geqslant}
\newcommand{\bbN}{\mathbb{N}}
\newcommand{\bbZ}{\mathbb{Z}}
\newcommand{\bbR}{\mathbb{R}}
\newcommand{\mcA}{\mathcal{A}}
\newcommand{\mcP}{\mathcal{P}}
\newcommand{\mcS}{\mathcal{S}}
\newcommand{\mcC}{\mathcal{C}}
\newcommand{\mcCs}{\mathcal{C}_\sigma}

\DeclareMathOperator{\supp}{supp}
\DeclareMathOperator{\patt}{patt}
\newcommand{\sbase}{\sigma_\textup{base}}
\newcommand{\srule}{\sigma_\textup{rule}}
\newcommand{\ws}{{\omega_\sigma}}
\newcommand{\Ssurf}{{\mathcal{S}_\textup{surf}}}
\newcommand{\Psurf}{{\mathcal{P}_\textup{surf}}}
\newcommand{\svectii}[2]{\big(\begin{smallmatrix} #1 \\ #2       \end{smallmatrix}\big)}

\newcommand{\myvcenter}[1]{\ensuremath{\vcenter{\hbox{#1}}}}

\newcommand{\tileB}[3]{
    \draw[draw=black, shift={(#1, #2)}] (0,0) -- (0,1) -- (1,1) -- (1,0) -- cycle;
    \node[shift={(#1, #2)}] at (0.5,0.5) {\small #3};
}
\newcommand{\tileR}[3]{
    \draw[-stealth, draw=black, shift={(#1, #2)}] (0,0.5) -- (1,0.5);
    \draw[draw=black, shift={(#1, #2)}] (0,0) -- (0,1) -- (1,1) -- (1,0) -- cycle;
    \node[shift={(#1, #2)}] at (0.5,0.25) {\scriptsize #3};
}
\newcommand{\tileU}[3]{
    \draw[-stealth, draw=black, shift={(#1, #2)}] (0.5,0) -- (0.5,1);
    \draw[draw=black, shift={(#1, #2)}] (0,0) -- (0,1) -- (1,1) -- (1,0) -- cycle;
    \node[shift={(#1, #2)}] at (0.25,0.5) {\scriptsize #3};
}
\newcommand{\tileL}[3]{
    \draw[-stealth, draw=black, shift={(#1, #2)}] (1,0.5) -- (0,0.5);
    \draw[draw=black, shift={(#1, #2)}] (0,0) -- (0,1) -- (1,1) -- (1,0) -- cycle;
    \node[shift={(#1, #2)}] at (0.5,0.25) {\scriptsize #3};
}
\newcommand{\tileD}[3]{
    \draw[-stealth, draw=black, shift={(#1, #2)}] (0.5,1) -- (0.5,0);
    \draw[draw=black, shift={(#1, #2)}] (0,0) -- (0,1) -- (1,1) -- (1,0) -- cycle;
    \node[shift={(#1, #2)}] at (0.25,0.5) {\scriptsize #3};
}

\newcommand{\tileBcol}[4]{
    \draw[draw=black, fill=black!#4, shift={(#1, #2)}] (0,0) -- (0,1) -- (1,1) -- (1,0) -- cycle;
    \node[shift={(#1, #2)}] at (0.5,0.5) {\small #3};
}

\title{\textbf{Consistency of multidimensional combinatorial substitutions}}

\author[1,2]{Timo Jolivet}
\author[1]{Jarkko Kari}

\affil[1]{
    FUNDIM,
    Department of Mathematics,
    University of Turku, Finland
}
\affil[2]{
    LIAFA,
    Universit\'e Paris 7, France
}

\date{}

\begin{document}

\maketitle

\begin{abstract}
Multidimensional combinatorial substitutions are
rules that replace symbols by finite patterns of symbols in $\bbZ^d$.
We focus on the case where the patterns are not necessarily rectangular,
which requires a specific description of the way they are glued together
in the image by a substitution.
Two problems can arise when defining a substitution in such a way:
it can fail to be consistent, and the patterns in an image by the substitution might overlap.

We prove that it is undecidable whether a two-dimensional substitution is consistent or overlapping,
and we provide practical algorithms to decide these properties in some particular cases.
\end{abstract}

\section{Introduction}
One-dimensional substitutions are a classical object of combinatorics on words.
An example is the map $\sigma : \{1,2,3\}^* \rightarrow \{1,2,3\}^*$ defined by
$\sigma(1) = 12$, $\sigma(2) = 13$ and $\sigma(3) = 1$.
The image by $\sigma$ of a word is easy to define, by \emph{concatenation}:
$\sigma(uv) = \sigma(u)\sigma(v)$ for all $u, v \in \{1,2,3\}^*$.
For example, $\sigma(1321) = 1211312$.
See \cite{PF02} for a detailed survey.

A natural generalization of substitutions to higher dimensions
is the case where the images of the letters
are squares of the same size, as for example in the
two-dimensional Thue-Morse substitution defined by
$1 \mapsto \begin{smallmatrix}1 & 2 \\ 2 & 1 \end{smallmatrix}$,
$2 \mapsto \begin{smallmatrix}2 & 1 \\ 1 & 2 \end{smallmatrix}$,
which can be iterated naturally, as shown below.
\begin{center}
$
1
\mapsto
\begin{smallmatrix}1 & 2 \\ 2 & 1 \end{smallmatrix}
\mapsto
\begin{smallmatrix}1 & 2 & 2 & 1 \\ 2 & 1 & 1 & 2 \\ 2 & 1 & 1 & 2 \\ 1 & 2 & 2 & 1\end{smallmatrix}
\mapsto
\begin{smallmatrix}
1 & 2 & 2 & 1 & 2 & 1 & 1 & 2 \\
2 & 1 & 1 & 2 & 1 & 2 & 2 & 1 \\
2 & 1 & 1 & 2 & 1 & 2 & 2 & 1 \\
1 & 2 & 2 & 1 & 2 & 1 & 1 & 2 \\
1 & 2 & 2 & 1 & 2 & 1 & 1 & 2 \\
2 & 1 & 1 & 2 & 1 & 2 & 2 & 1 \\
2 & 1 & 1 & 2 & 1 & 2 & 2 & 1 \\
1 & 2 & 2 & 1 & 2 & 1 & 1 & 2
\end{smallmatrix}
$
\end{center}
Similar generalizations are possible with rectangular shapes
that have compatible edge lengths;
see the survey \cite{Fra08}.

We are interested in the more general case
where the images of the letters have arbitrary shapes (not necessarily rectangular),
which are \emph{a priori} not compatible with concatenation.
For example, if $\sigma$ is a substitution defined by:
\begin{center}
\myvcenter{\begin{tikzpicture}[x=2.5mm, y=2.5mm]
\draw[black]
(0, 1) -- (0, 0) -- (1, 0) -- (1, 1) -- cycle;
\node at (0.5,0.5) {\scriptsize $1$};
\end{tikzpicture}}
\ $\mapsto$ \
\myvcenter{\begin{tikzpicture}[x=2.5mm, y=2.5mm]
\draw[black]
(-1, 1) -- (-1, 0) -- (0, 0) -- (0, -1) -- (1, -1) -- (1, 0) -- (1, 1) -- (1, 2) -- (0, 2) -- (0, 1) -- cycle;
\node at (0.5,0.5) {\scriptsize $1$};
\node at (0.5,1.5) {\scriptsize $2$};
\node at (0.5,-0.5) {\scriptsize $1$};
\node at (-0.5,0.5) {\scriptsize $3$};
\end{tikzpicture}}
  \qquad
\myvcenter{\begin{tikzpicture}[x=2.5mm, y=2.5mm]
\draw[black]
(0, 1) -- (0, 0) -- (1, 0) -- (1, 1) -- cycle;
\node at (0.5,0.5) {\scriptsize $2$};
\end{tikzpicture}}
\ $\mapsto$ \
\myvcenter{\begin{tikzpicture}[x=2.5mm, y=2.5mm]
\draw[black]
(0, 1) -- (0, 0) -- (1, 0) -- (1, 1) -- (1, 2) -- (0, 2) -- cycle;
\node at (0.5,0.5) {\scriptsize $1$};
\node at (0.5,1.5) {\scriptsize $2$};
\end{tikzpicture}}
  \qquad
\myvcenter{\begin{tikzpicture}[x=2.5mm, y=2.5mm]
\draw[black]
(0, 1) -- (0, 0) -- (1, 0) -- (1, 1) -- cycle;
\node at (0.5,0.5) {\scriptsize $3$};
\end{tikzpicture}}
\ $\mapsto$ \
\myvcenter{\begin{tikzpicture}[x=2.5mm, y=2.5mm]
\draw[black]
(-1, 1) -- (-1, 0) -- (0, 0) -- (1, 0) -- (1, 1) -- (1, 2) -- (0, 2) -- (0, 1) -- cycle;
\node at (0.5,0.5) {\scriptsize $1$};
\node at (0.5,1.5) {\scriptsize $2$};
\node at (-0.5,0.5) {\scriptsize $3$};
\end{tikzpicture}},
\end{center}
how can we define the image by $\sigma$ of
patterns that consist of more than one cell?
A natural approach is to give explicit \emph{concatenation rules}, such as:
\begin{center}
\myvcenter{\begin{tikzpicture}[x=2.5mm, y=2.5mm]
\draw[black]
(0, 1) -- (0, 0) -- (1, 0) -- (1, 1) -- cycle;
\node at (0.5,0.5) {\scriptsize $1$};
\definecolor{cellcolor}{rgb}{0.65,0.65,0.65}
\fill[fill=cellcolor, draw=black]
(0, 2) -- (0, 1) -- (1, 1) -- (1, 2) -- cycle;
\node at (0.5,1.5) {\scriptsize $2$};
\end{tikzpicture}}
\ $\mapsto$ \
\myvcenter{\begin{tikzpicture}[x=2.5mm, y=2.5mm]
\draw[black]
(-1, 1) -- (-1, 0) -- (0, 0) -- (0, -1) -- (1, -1) -- (1, 0) -- (1, 1) -- (1, 2) -- (0, 2) -- (0, 1) -- cycle;
\node at (0.5,0.5) {\scriptsize $1$};
\node at (0.5,1.5) {\scriptsize $2$};
\node at (-0.5,0.5) {\scriptsize $3$};
\node at (0.5,-0.5) {\scriptsize $1$};
\definecolor{cellcolor}{rgb}{0.65,0.65,0.65}
\fill[fill=cellcolor, draw=black]
(1, 2) -- (1, 1) -- (2, 1) -- (2, 2) -- (2, 3) -- (1, 3) -- cycle;
\node at (1.5,1.5) {\scriptsize $1$};
\node at (1.5,2.5) {\scriptsize $2$};
\end{tikzpicture}}
 \quad
\myvcenter{\begin{tikzpicture}[x=2.5mm, y=2.5mm]
\draw[black]
(0, 1) -- (0, 0) -- (1, 0) -- (1, 1) -- cycle;
\node at (0.5,0.5) {\scriptsize $3$};
\definecolor{cellcolor}{rgb}{0.65,0.65,0.65}
\fill[fill=cellcolor, draw=black]
(1, 1) -- (1, 0) -- (2, 0) -- (2, 1) -- cycle;
\node at (1.5,0.5) {\scriptsize $1$};
\end{tikzpicture}}
\ $\mapsto$ \
\myvcenter{\begin{tikzpicture}[x=2.5mm, y=2.5mm]
\draw[black]
(-1, 2) -- (0, 2) -- (0, 3) -- (0, 4) -- (-1, 4) -- (-1, 3) -- (-2, 3) -- (-2, 2) -- cycle;
\node at (-0.5,2.5) {\scriptsize $1$};
\node at (-0.5,3.5) {\scriptsize $2$};
\node at (-1.5,2.5) {\scriptsize $3$};
\definecolor{cellcolor}{rgb}{0.65,0.65,0.65}
\fill[fill=cellcolor, draw=black]
(-1, 2) -- (-1, 1) -- (0, 1) -- (0, 0) -- (1, 0) -- (1, 1) -- (1, 2) -- (1, 3) -- (0, 3) -- (0, 2) -- cycle;
\node at (0.5,0.5) {\scriptsize $1$};
\node at (0.5,1.5) {\scriptsize $1$};
\node at (0.5,2.5) {\scriptsize $2$};
\node at (-0.5,1.5) {\scriptsize $3$};
\end{tikzpicture}}
 \quad
\myvcenter{\begin{tikzpicture}[x=2.5mm, y=2.5mm]
\draw[black]
(0, 1) -- (0, 0) -- (1, 0) -- (1, 1) -- cycle;
\node at (0.5,0.5) {\scriptsize $1$};
\definecolor{cellcolor}{rgb}{0.65,0.65,0.65}
\fill[fill=cellcolor, draw=black]
(0, 2) -- (0, 1) -- (1, 1) -- (1, 2) -- cycle;
\node at (0.5,1.5) {\scriptsize $1$};
\end{tikzpicture}}
\ $\mapsto$
\myvcenter{\begin{tikzpicture}[x=2.5mm, y=2.5mm]
\draw[black]
(-1, 1) -- (-1, 0) -- (0, 0) -- (0, -1) -- (1, -1) -- (1, 0) -- (1, 1) -- (1, 2) -- (0, 2) -- (0, 1) -- cycle;
\node at (0.5,0.5) {\scriptsize $1$};
\node at (0.5,1.5) {\scriptsize $2$};
\node at (0.5,-0.5) {\scriptsize $1$};
\node at (-0.5,0.5) {\scriptsize $3$};
\definecolor{cellcolor}{rgb}{0.65,0.65,0.65}
\fill[fill=cellcolor, draw=black]
(0, 3) -- (0, 2) -- (1, 2) -- (1, 1) -- (2, 1) -- (2, 2) -- (2, 3) -- (2, 4) -- (1, 4) -- (1, 3) -- cycle;
\node at (1.5,1.5) {\scriptsize $1$};
\node at (1.5,2.5) {\scriptsize $1$};
\node at (1.5,3.5) {\scriptsize $2$};
\node at (0.5,2.5) {\scriptsize $3$};
\end{tikzpicture}}
 \quad
\myvcenter{\begin{tikzpicture}[x=2.5mm, y=2.5mm]
\draw[black]
(0, 1) -- (0, 0) -- (1, 0) -- (1, 1) -- cycle;
\node at (0.5,0.5) {\scriptsize $2$};
\definecolor{cellcolor}{rgb}{0.65,0.65,0.65}
\fill[fill=cellcolor, draw=black]
(1, 1) -- (1, 0) -- (2, 0) -- (2, 1) -- cycle;
\node at (1.5,0.5) {\scriptsize $1$};
\end{tikzpicture}}
\ $\mapsto$ \
\myvcenter{\begin{tikzpicture}[x=2.5mm, y=2.5mm]
\draw[black]
(-1, 2) -- (-1, 1) -- (0, 1) -- (0, 2) -- (0, 3) -- (-1, 3) -- cycle;
\node at (-0.5,1.5) {\scriptsize $1$};
\node at (-0.5,2.5) {\scriptsize $2$};
\definecolor{cellcolor}{rgb}{0.65,0.65,0.65}
\fill[fill=cellcolor, draw=black]
(-1, 1) -- (-1, 0) -- (0, 0) -- (0, -1) -- (1, -1) -- (1, 0) -- (1, 1) -- (1, 2) -- (0, 2) -- (0, 1) -- cycle;
\node at (0.5,0.5) {\scriptsize $1$};
\node at (0.5,1.5) {\scriptsize $2$};
\node at (0.5,-0.5) {\scriptsize $1$};
\node at (-0.5,0.5) {\scriptsize $3$};
\end{tikzpicture}}
 \quad
\myvcenter{\begin{tikzpicture}[x=2.5mm, y=2.5mm]
\draw[black]
(0, 1) -- (0, 0) -- (1, 0) -- (1, 1) -- cycle;
\node at (0.5,0.5) {\scriptsize $3$};
\definecolor{cellcolor}{rgb}{0.65,0.65,0.65}
\fill[fill=cellcolor, draw=black]
(0, 2) -- (0, 1) -- (1, 1) -- (1, 2) -- cycle;
\node at (0.5,1.5) {\scriptsize $1$};
\end{tikzpicture}}
\ $\mapsto$
\myvcenter{\begin{tikzpicture}[x=2.5mm, y=2.5mm]
\draw[black]
(-1, 0) -- (-1, -1) -- (-2, -1) -- (-2, -2) -- (-1, -2) -- (0, -2) -- (0, -1) -- (0, 0) -- cycle;
\node at (-0.5,-1.5) {\scriptsize $1$};
\node at (-0.5,-0.5) {\scriptsize $2$};
\node at (-1.5,-1.5) {\scriptsize $3$};
\definecolor{cellcolor}{rgb}{0.65,0.65,0.65}
\fill[fill=cellcolor, draw=black]
(-1, 1) -- (-1, 0) -- (0, 0) -- (0, -1) -- (1, -1) -- (1, 0) -- (1, 1) -- (1, 2) -- (0, 2) -- (0, 1) -- cycle;
\node at (0.5,0.5) {\scriptsize $1$};
\node at (0.5,1.5) {\scriptsize $2$};
\node at (-0.5,0.5) {\scriptsize $3$};
\node at (0.5,-0.5) {\scriptsize $1$};
\end{tikzpicture}}.
\end{center}
Using these rules, we can compute the image of every pattern that can be ``covered''
by the two-cell patterns on the left-hand sides of the rules, as shown below.
\begin{center}
\myvcenter{%
\begin{tikzpicture}[x=2.5mm, y=2.5mm]
\definecolor{cellcolor}{rgb}{1,1,1}
\fill[fill=cellcolor, draw=black]
(0, 1) -- (0, 0) -- (1, 0) -- (1, 1) -- cycle;
\node at (0.5,0.5) {\scriptsize $3$};
\definecolor{cellcolor}{rgb}{0.8,0.8,0.8}
\fill[fill=cellcolor, draw=black]
(1, 1) -- (1, 0) -- (2, 0) -- (2, 1) -- cycle;
\definecolor{cellcolor}{rgb}{0.4,0.4,0.4}
\fill[fill=cellcolor, draw=black,shift={(1,1)}]
(1, 1) -- (1, 0) -- (2, 0) -- (2, 1) -- cycle;
\node at (1.5,0.5) {\scriptsize $1$};
\node at (2.5,1.5) {\scriptsize $1$};
\definecolor{cellcolor}{rgb}{0.6,0.6,0.6}
\fill[fill=cellcolor, draw=black]
(1, 2) -- (1, 1) -- (2, 1) -- (2, 2) -- cycle;
\node at (1.5,1.5) {\scriptsize $2$};
\end{tikzpicture}}
$\mapsto$
\myvcenter{%
\begin{tikzpicture}[x=2.5mm, y=2.5mm]
\definecolor{cellcolor}{rgb}{1,1,1}
\fill[fill=cellcolor, draw=black]
(-1, 1) -- (-1, 0) -- (0, 0) -- (1, 0) -- (1, 1) -- (1, 2) -- (0, 2) -- (0, 1) -- cycle;
\node at (0.5,0.5) {\scriptsize $1$};
\node at (0.5,1.5) {\scriptsize $2$};
\node at (-0.5,0.5) {\scriptsize $3$};
\definecolor{cellcolor}{rgb}{0.8,0.8,0.8}
\fill[fill=cellcolor, draw=black]
(0, 0) -- (0, -1) -- (1, -1) -- (1, -2) -- (2, -2) -- (2, -1) -- (2, 0) -- (2, 1) -- (1, 1) -- (1, 0) -- cycle;
\definecolor{cellcolor}{rgb}{0.4,0.4,0.4}
\fill[fill=cellcolor, draw=black, shift={(2,0)}]
(0, 0) -- (0, -1) -- (1, -1) -- (1, -2) -- (2, -2) -- (2, -1) -- (2, 0) -- (2, 1) -- (1, 1) -- (1, 0) -- cycle;
\node at (0.5,-0.5) {\scriptsize $3$};
\node at (1.5,-0.5) {\scriptsize $1$};
\node at (1.5,0.5) {\scriptsize $2$};
\node at (1.5,-1.5) {\scriptsize $1$};
\node at (2.5,-0.5) {\scriptsize $3$};
\node at (3.5,-0.5) {\scriptsize $1$};
\node at (3.5,0.5) {\scriptsize $2$};
\node at (3.5,-1.5) {\scriptsize $1$};
\definecolor{cellcolor}{rgb}{0.6,0.6,0.6}
\fill[fill=cellcolor, draw=black]
(2, 1) -- (2, 0) -- (3, 0) -- (3, 1) -- (3, 2) -- (2, 2) -- cycle;
\node at (2.5,0.5) {\scriptsize $1$};
\node at (2.5,1.5) {\scriptsize $2$};
\end{tikzpicture}}
\end{center}

In this paper we study the two problems
that can arise when multidimensional concatenation is defined as above:
\begin{enumerate}
\itemsep=0pt \parskip=0pt
\item
The resulting substitution can fail to be \emph{consistent}:
depending on the sequence of concatenation rules that are used,
a pattern might have two different images.
\item
The resulting substitution can fail to be \emph{non-overlapping}:
the images of the cells of a pattern might overlap.
\end{enumerate}
The substitutions which are consistent and non-overlapping correspond to ``well defined'' substitutions,
and we would like to be able to detect them algorithmically.
We will prove that consistency and non-overlapping are undecidable properties
for two-dimensional combinatorial substitutions (Theorems~\ref{theo:undcons} and~\ref{theo:undnono}),
but that these properties are decidable in the case of \emph{domino-complete} substitutions,
that is, when the concatenation rules are given for every possible domino
(Theorems~\ref{theo:deccons} and~\ref{theo:decnono}).
This answers the decidability question raised in \cite{Fer07}.

As an application of the methods developed for the decidability results,
we provide combinatorial proofs of the consistency and non-overlapping
of some particular two-dimensional substitutions,
using a slightly more general definition of domino-completeness
(Theorem~\ref{theo:decconsmieux}, Section~\ref{sec:applis}).
Such proofs have been requested in \cite{ABS04, Fra08}.

Combinatorial substitutions as defined in this article were introduced in \cite{ABS04} under the name ``local rules'',
to study a particular example in the context of substitutions of unit faces of discrete planes.
They were defined in generality in \cite{Fer07},
and are related to the substitutions found in \cite{Fra03}
defined using the dual graph of a pattern.
See \cite{Fra08} for a survey about multidimensional substitutions.

It is important to remark that the properties studied in this paper (consistency and overlapping)
only concern a \emph{single} iteration of the substitution.
This is a necessary first step for the further study of
properties related with iterating a substitution indefinitely on a given pattern
(see the last section and the related open problems).

\section{Definitions}
\label{sec:def}

\subsection{Cells and patterns}
Let $\mcA$ denote a set of symbols.
A \emph{$d$-dimensional cell} is a couple $c = [v, t]$,
where $v \in \bbZ^d$ is the \emph{vector} of $c$
and $t \in \mcA$ is the \emph{type} of $c$.
A \emph{$d$-dimensional pattern} is a finite union of $d$-dimensional cells with distinct vectors.
Translation $P + v$ of a pattern $P$ by $v \in \bbZ^d$ is defined in the natural way.
The \emph{support} of a pattern is $\supp(P) = \{v : [v, t] \in P\}$.

Many of the substitutions we will encounter later use \emph{dominoes},
which are two-dimensional patterns that consists of two cells of vectors $v$ and $v'$
such that $v' - v \in \{(\pm 1, 0), (0, \pm 1))\}$.

\subsection{Substitutions}
A \emph{$d$-dimensional substitution} $\sigma$ on alphabet $\mcA$ is defined by:
\begin{itemize}
\itemsep=0pt \parskip=0pt
\item
    a \emph{base rule}:
    an application $\sbase$ from $\mcA$ to the set of $d$-dimensional patterns,
\item
    a finite set of \emph{concatenation rules} $(t, t', u) \mapsto v$,
    where $t, t' \in \mcA$ and $u, v \in \bbZ^d$.
\end{itemize}
The way to interpret this definition is the following:
$\sbase$ replaces each cell of a pattern by a pattern,
and the concatenation rules describe how to place the images of the cells relatively to each other.
The intuitive meaning of ``$(t, t', u) \mapsto v$'' is:
two cells of types $t$ and $t'$ separated by $u$
must be mapped by $\sigma$ to
the two patterns $\sbase(t)$ and $\sbase(t')$ separated by $v$.
(A precise definition is given below.)
From now on we only consider \emph{deterministic} substitutions,
which means that if a rule has a left-hand side $(t,t',u)$,
then there is no other rule with left-hand side either $(t,t',u)$ or $(t',t,-u)$.

We will need the following notation:
\begin{itemize}
\itemsep=0pt \parskip=0pt
\item
We extend $\sbase$ from $\mcA$ to the set of cells naturally:
$\sbase(c) = \sbase(t)$ for a cell $c = [v,t]$.
(Only the type of $c$ is taken into account by $\sbase$.)
\item
The set of the \emph{starting patterns} of $\sigma$ is
\[
\mcCs = \{\{[0, t], [u, t']\} : (t, t', u) \mapsto v \text{ is a rule of } \sigma \text{ for some } v\}.
\]
It corresponds to the patterns at the left-hand sides of the concatenation rules of $\sigma$.
(This set contains only patterns of size two.)
\item
If $c = [u, t]$ and $c' = [u',t']$ are cells, we denote
\[
\srule(c, c') \ = \ \left \{
\begin{array}{cl}
v & \text{ if } (t, t', u'-u) \mapsto v \text{ is a rule of } \sigma \\
-v & \text{ if } (t', t, u-u') \mapsto v \text{ is a rule of } \sigma
\end{array}
\right.,
\]
and $\srule$ is not defined otherwise.
\end{itemize}

A \emph{domino substitution} is a two-dimensional substitution such that
for every rule $(t, t', u) \mapsto v$,
we have $u \in \{\pm (1,0), \pm (0,1)\}$.
A \emph{domino-to-domino} substitution is a domino substitution such that
for every rule $(t, t', u) \mapsto v$,
we have $v \in \{\pm (1,0), \pm (0,1)\}$,
and the patterns $\sbase(t)$ and $\sbase(t')$ both consist of a single cell of vector $(0,0)$.

\begin{exam}
The combinatorial substitution given in the introduction is formally defined as follows:
it is the two-dimensional substitution defined on the alphabet $\{1,2,3\}$
with the following base rule (on the left) and concatenation rules (on the right).
\[
\begin{array}{rclcrcl}
& & & & (1, 2, (0,1)) & \mapsto & (1,2) \\
1 & \mapsto & \{[(0, 0), 1], [(0, 1), 1], [(0, 2), 2], [(-1, 1), 3]\} & &
  (3, 1, (1,0)) & \mapsto & (2,-2) \\
2 & \mapsto & \{[(0, 0), 1], [(0, 1), 2]\} & &
  (1, 1, (0,1)) & \mapsto & (1,2) \\
3 & \mapsto & \{[(0, 0), 3], [(1, 0), 1], [(1, 1), 2]\} & &
  (2, 1, (1,0)) & \mapsto & (1,-2) \\
& & & & (3, 1, (0,1)) & \mapsto & (2,1).
\end{array}
\]
\end{exam}

\subsection{Paths, covers, image vectors}

Let $\mcC$ be a finite set of patterns each consisting of two cells.
A \emph{$\mcC$-path from $c_1$ to $c_n$} is a finite sequence of cells $\gamma = (c_1, \ldots, c_n)$ such that
$\{c_i, c_{i+1}\}$ is a translated copy of an element of $\mcC$ for all $1 \leq i \leq n-1$,
and such that $c_i = c_j$ if $c_i$ and $c_j$ have the same vector.
(Paths are hence allowed to self-overlap, but the overlapping cells must agree.)
If $c_1 = c_n$, then $\gamma$ is called a \emph{$\mcC$-loop}.
A path (or a loop) is \emph{simple} if it does not self-intersect.
If all the cells of $\gamma$ are contained in $P$
then it is called a \emph{$\mcC$-path of $P$}.

A pattern $P$ is \emph{$\mcC$-covered}
\index{$\mcC$-covering}
\index{covering!$\mcC$-covering}
if for every $c, c' \in P$,
there exists a $\mcC$-path $\gamma$ of $P$ from $c$ to $c'$.
Let $\sigma$ be a substitution and $\gamma = (c_1, \ldots, c_n)$ be a $\mcCs$-path.
We denote by $\ws(\gamma)$ the \emph{image vector of $\gamma$} defined by
\[
\ws(\gamma) \ = \ \sum_{i=1}^{n-1} \srule(c_i, c_{i+1}).
\]

\subsection{Consistency and non-overlapping}
\label{sec:consnono}
Let $\sigma$ be a substitution and $P$ be a $\mcCs$-covered pattern.
We say that $\sigma$ is:
\begin{itemize}
\itemsep=0pt \parskip=0pt
\item
\emph{consistent on $P$} if
for every cells $c, c' \in P$ and for every $\mcCs$-paths $\gamma, \gamma'$ of $P$ from $c$ to $c'$,
we have $\ws(\gamma) = \ws(\gamma')$,
\emph{i.e.},
if the placement of the images does not depend on the path used.
\item
\emph{non-overlapping on $P$} if
for every cells $c, c' \in P$ such that $c \neq c'$
and for every $\mcCs$-path $\gamma$ of $P$ from $c$ to $c'$,
we have
$\supp(\sbase(c)) \ \cap \ (\supp(\ws(\gamma) + \sbase(c'))) \ = \ \varnothing$,
\emph{i.e.},
if two distinct cells have non-overlapping images.
\end{itemize}
If $\sigma$ is consistent on every $\mcCs$-covered pattern,
then $\sigma$ is said to be \emph{consistent}.
(The same goes for non-overlapping.)
Examples of inconsistent and overlapping substitutions
will be given in Examples~\ref{exam:inconsistent} and~\ref{exam:overlapping}.

\begin{prop}
\label{prop:loopchar}
Let $\sigma$ be a substitution and $P$ be a pattern.
The following statements are equivalent.
\begin{enumerate}
\itemsep=0pt \parskip=0pt
\item
\label{item:loopchar1}
$\sigma$ is consistent on $P$.
\item
\label{item:loopchar2}
For every $\mcCs$-loop $\gamma$ of $P$, we have $\ws(\gamma) = 0$.
\item
\label{item:loopchar3}
For every simple $\mcCs$-loop $\gamma$ of $P$, we have $\ws(\gamma) = 0$.
\end{enumerate}
\end{prop}

\begin{proof}
(\ref{item:loopchar1}) $\Leftrightarrow$ (\ref{item:loopchar2}).
Suppose that $\sigma$ is consistent,
and let $\gamma = (c_1, \ldots, c_n)$ be a $\mcCs$-loop of $P$.
Let $\gamma' = (c_1, \ldots, c_{n-1})$
and $\gamma'' =(c_n, c_{n-1})$.
Because $c_n = c_1$ and $\sigma$ is consistent
we have $\ws(\gamma') = \ws(\gamma'')$, so
\[
\begin{array}{rcl}
\ws(\gamma)
    & \ = \ & \displaystyle \sum_{i=1}^{n-1} \srule(c_i, c_{i+1})
      \ = \ \displaystyle \sum_{i=1}^{n-2} \srule(c_i, c_{i+1}) - \srule(c_n, c_{n-1}) \\
    & \ = \ & \ws(\gamma') - \ws(\gamma'')
      \ = \  0.
\end{array}
\]
Conversely, let $\gamma = (c_1, \ldots, c_n)$ and $\gamma' = (c'_1, \ldots c'_m)$
be two $\mcCs$-paths of $P$ with $c_1 = c'_1$ and $c_n = c'_m$.
Now, $\gamma'' \coloneqq (c_1, \ldots, c_{n-1}, c_n, c'_{m-1}, \ldots, c'_1)$ is a $\mcCs$-loop,
so $\ws(\gamma'') = 0$, which implies that
$\ws(\gamma) \ = \ \ws(\gamma) - \ws(\gamma'') \ = \ \ws(\gamma) - (\ws(\gamma) - \ws(\gamma')) \ = \ \ws(\gamma')$.

(\ref{item:loopchar2}) $\Leftrightarrow$ (\ref{item:loopchar3}).
Implication ``$\Rightarrow$'' is trivial.
For the converse, suppose that there exists a non-simple $\mcCs$-loop $\gamma = (c_1, \ldots, c_n)$ of $P$
such that $\ws(\gamma) \neq 0$,
and let $i < j < n$ such that $c_i = c_j$.
Let $\gamma' = (c_i, \ldots, c_j)$
and $\gamma'' = (c_1, \ldots, c_i, c_{j+1}, \ldots, c_n)$.
We have $\ws(\gamma') + \ws(\gamma'') = \ws(\gamma) \neq 0$,
so $\ws(\gamma') \neq 0$ or $\ws(\gamma'') \neq 0$.
Repeating this operation inductively yields the existence of
a simple loop (strictly smaller than $\gamma$) which does not overlap itself
and which has a nonzero image vector.
(The loop cannot reduce to a single cell because we assumed that $\ws(\gamma) \neq 0$.)
\end{proof}

\subsection{Image by a substitution}
Let $\sigma$ be a non-overlapping substitution.
Let $P$ be a $\mcCs$-covered pattern and $c_0$ be a cell of $P$.
An \emph{image of $P$ by $\sigma$ computed from $c_0$} is a pattern
\[
\bigcup_{c \in P} (\sbase(c) \ + \ \ws(\gamma_c)),
\]
where for each $c \in P$, $\gamma_c$ is a $\mcCs$-path from $c_0$ to $c$.
This union of patterns is indeed a pattern
because the cells have distinct positions ($\sigma$ is non-overlapping).

If $\sigma$ is consistent, this pattern is uniquely defined
because it does not depend on the choice of the paths $\gamma_c$, by consistency of $\sigma$.
In this case, the image of $P$ by $\sigma$ computed from $c_0$
is denoted by $\sigma(P,c_0)$, but because $c_0$ only affects by a translation
we will use the simpler notation $\sigma(P)$ when the translation is irrelevant.

To explicitly compute the image of an $\mcCs$-covered pattern $P$,
we start by constructing a tree whose vertices are the cells of $P$,
where two cells are connected if they both belong to a same pattern of $\mcCs$.
(Such a connected tree exists because $P$ is $\mcCs$-covered.)
We then choose a ``root cell'' ($c_0$ in the definition above),
and we construct the image of $P$ incrementally,
starting from $c_0$ and following the edges in the tree.
This is shown in Example~\ref{exam:JP} below.

\begin{exam}
\label{exam:JP}
Let $\sigma$ be the two-dimensional substitution defined on the alphabet $\{1,2,3\}$
with the base rule
\[
\begin{array}{rclcl}
1 & \mapsto & \{[(0, 0), 2]\} & = &
\myvcenter{\begin{tikzpicture}[x=2.5mm, y=2.5mm]
\draw[black]
(0, 1) -- (0, 0) -- (1, 0) -- (1, 1) -- cycle;
\node at (0.5,0.5) {\scriptsize $2$};
\end{tikzpicture}} \\
2 & \mapsto & \{[(0, 0), 3]\} & = &
\myvcenter{\begin{tikzpicture}[x=2.5mm, y=2.5mm]
\draw[black]
(0, 1) -- (0, 0) -- (1, 0) -- (1, 1) -- cycle;
\node at (0.5,0.5) {\scriptsize $3$};
\end{tikzpicture}} \\
3 & \mapsto & \{[(0, 0), 1], [(1, 0), 3]\} & = &
\myvcenter{\begin{tikzpicture}[x=2.5mm, y=2.5mm]
\draw[black]
(0, 1) -- (0, 0) -- (1, 0) -- (2, 0) -- (2, 1) -- (1, 1) -- cycle;
\node at (0.5,0.5) {\scriptsize $1$};
\node at (1.5,0.5) {\scriptsize $3$};
\end{tikzpicture}}
\end{array}
\]
and the concatenation rules
\[
\begin{array}{llll}
(2, 3, \svectii{1}{1}) \mapsto \svectii{-2}{0} &
(3, 2, \svectii{0}{1}) \mapsto \svectii{-1}{-1} &
(1, 3, \svectii{1}{0}) \mapsto \svectii{0}{1} &
(3, 3, \svectii{1}{0}) \mapsto \svectii{0}{1} \\
(3, 1, \svectii{1}{1}) \mapsto \svectii{-1}{0} &
(3, 3, \svectii{0}{1}) \mapsto \svectii{-2}{-1} &
(2, 1, \svectii{1}{0}) \mapsto \svectii{0}{1}
\end{array}
\]
which can also be represented as follows
\begin{center}
$
\begin{array}{rclcrclcrclcrcl}
\myvcenter{\begin{tikzpicture}[x=2.5mm, y=2.5mm]
\draw[black]
(0, 1) -- (0, 0) -- (1, 0) -- (1, 1) -- cycle;
\node at (0.5,0.5) {\scriptsize $2$};
\definecolor{cellcolor}{rgb}{0.65,0.65,0.65}
\fill[fill=cellcolor, draw=black]
(1, 2) -- (1, 1) -- (2, 1) -- (2, 2) -- cycle;
\node at (1.5,1.5) {\scriptsize $3$};
\end{tikzpicture}}
& \mapsto &
\myvcenter{\begin{tikzpicture}[x=2.5mm, y=2.5mm]
\draw[black]
(0, 1) -- (0, 0) -- (1, 0) -- (1, 1) -- cycle;
\node at (0.5,0.5) {\scriptsize $3$};
\definecolor{cellcolor}{rgb}{0.65,0.65,0.65}
\fill[fill=cellcolor, draw=black]
(-1, 0) -- (0, 0) -- (0, 1) -- (-1, 1) -- (-2, 1) -- (-2, 0) -- cycle;
\node at (-1.5,0.5) {\scriptsize $1$};
\node at (-0.5,0.5) {\scriptsize $3$};
\end{tikzpicture}}
& \quad &
\myvcenter{\begin{tikzpicture}[x=2.5mm, y=2.5mm]
\draw[black]
(0, 1) -- (0, 0) -- (1, 0) -- (1, 1) -- cycle;
\node at (0.5,0.5) {\scriptsize $2$};
\definecolor{cellcolor}{rgb}{0.65,0.65,0.65}
\fill[fill=cellcolor, draw=black]
(0, 0) -- (0, -1) -- (1, -1) -- (1, 0) -- cycle;
\node at (0.5,-0.5) {\scriptsize $3$};
\end{tikzpicture}}
& \mapsto &
\myvcenter{\begin{tikzpicture}[x=2.5mm, y=2.5mm]
\draw[black]
(0, 1) -- (0, 0) -- (1, 0) -- (1, 1) -- cycle;
\node at (0.5,0.5) {\scriptsize $3$};
\definecolor{cellcolor}{rgb}{0.65,0.65,0.65}
\fill[fill=cellcolor, draw=black]
(1, 2) -- (1, 1) -- (2, 1) -- (3, 1) -- (3, 2) -- (2, 2) -- cycle;
\node at (1.5,1.5) {\scriptsize $1$};
\node at (2.5,1.5) {\scriptsize $3$};
\end{tikzpicture}}
& \quad &
\myvcenter{\begin{tikzpicture}[x=2.5mm, y=2.5mm]
\draw[black]
(0, 1) -- (0, 0) -- (1, 0) -- (1, 1) -- cycle;
\node at (0.5,0.5) {\scriptsize $1$};
\definecolor{cellcolor}{rgb}{0.65,0.65,0.65}
\fill[fill=cellcolor, draw=black]
(1, 1) -- (1, 0) -- (2, 0) -- (2, 1) -- cycle;
\node at (1.5,0.5) {\scriptsize $3$};
\end{tikzpicture}}
& \mapsto &
\myvcenter{\begin{tikzpicture}[x=2.5mm, y=2.5mm]
\draw[black]
(0, 0) -- (0, -1) -- (1, -1) -- (1, 0) -- cycle;
\node at (0.5,-0.5) {\scriptsize $2$};
\definecolor{cellcolor}{rgb}{0.65,0.65,0.65}
\fill[fill=cellcolor, draw=black]
(0, 1) -- (0, 0) -- (1, 0) -- (2, 0) -- (2, 1) -- (1, 1) -- cycle;
\node at (0.5,0.5) {\scriptsize $1$};
\node at (1.5,0.5) {\scriptsize $3$};
\end{tikzpicture}}
& \quad &
\myvcenter{\begin{tikzpicture}[x=2.5mm, y=2.5mm]
\draw[black]
(0, 1) -- (0, 0) -- (1, 0) -- (1, 1) -- cycle;
\node at (0.5,0.5) {\scriptsize $3$};
\definecolor{cellcolor}{rgb}{0.65,0.65,0.65}
\fill[fill=cellcolor, draw=black]
(1, 1) -- (1, 0) -- (2, 0) -- (2, 1) -- cycle;
\node at (1.5,0.5) {\scriptsize $3$};
\end{tikzpicture}}
& \mapsto &
\myvcenter{\begin{tikzpicture}[x=2.5mm, y=2.5mm]
\draw[black]
(0, 1) -- (0, 0) -- (1, 0) -- (2, 0) -- (2, 1) -- (1, 1) -- cycle;
\node at (0.5,0.5) {\scriptsize $1$};
\node at (1.5,0.5) {\scriptsize $3$};
\definecolor{cellcolor}{rgb}{0.65,0.65,0.65}
\fill[fill=cellcolor, draw=black]
(0, 2) -- (0, 1) -- (1, 1) -- (2, 1) -- (2, 2) -- (1, 2) -- cycle;
\node at (0.5,1.5) {\scriptsize $1$};
\node at (1.5,1.5) {\scriptsize $3$};
\end{tikzpicture}}
 \\
\end{array}
$
\end{center}
To compute the image of
\myvcenter{%
}
by $\sigma$, we can use the $\mcCs$-covering
\myvcenter{}:
\[
\begin{array}{rcl}
\sigma\big(\myvcenter{\begin{tikzpicture}[x=2.5mm, y=2.5mm]
\definecolor{cellcolor}{gray}{0.65}
\fill[fill=cellcolor, draw=black]
(0,0) -- (1,0) -- (1,1) -- (0,1) -- cycle;
\node at (0.5,0.5) {\scriptsize $2$};
\definecolor{cellcolor}{gray}{1}
\fill[fill=cellcolor, draw=black]
(0,1) -- (1,1) -- (1,2) -- (0,2) -- cycle;
\node at (0.5,1.5) {\scriptsize $1$};
\fill[fill=cellcolor, draw=black]
(1,0) -- (2,0) -- (2,1) -- (1,1) -- cycle;
\node at (1.5,0.5) {\scriptsize $1$};
\fill[fill=cellcolor, draw=black]
(2,0) -- (3,0) -- (3,1) -- (2,1) -- cycle;
\node at (2.5,0.5) {\scriptsize $3$};
\fill[fill=cellcolor, draw=black]
(1,1) -- (2,1) -- (2,2) -- (1,2) -- cycle;
\node at (1.5,1.5) {\scriptsize $3$};
\fill[fill=cellcolor, draw=black]
(2,1) -- (3,1) -- (3,2) -- (2,2) -- cycle;
\node at (2.5,1.5) {\scriptsize $3$};
\end{tikzpicture}}\big)
    & = & \{[0,0], 3\} \\
    & \cup & \{[(0, 0), 1], [(1, 0), 3]\} + (-2,0) \\
    & \cup & \{[(0, 0), 2]\} + (-2,0) + (0,-1) \\
    & \cup & \{[(0, 0), 1], [(1, 0), 3]\} + (-2,0) + (0,1) \\
    & \cup & \{[(0, 0), 1], [(1, 0), 3]\} + (-2,0) + (0,1) + (2,1)  \\
    & \cup & \{[(0, 0), 2]\} + (-2,0) + (0,1) + (2,1) + (0,-1) \\
    & = & \{[(0, 0), 3], [(-2, 0), 1], [(-1, 0), 3], [(-2, -1), 2], \\
    & & \phantom{\}} [(-2, 1), 1], [(-1, 1), 3], [(0, 1), 2], [(0, 2), 1], [(1, 2), 3]\} \\
\end{array}
\]
We chose to place the image of the cell
\myvcenter{%
}
first, at position $(0,0)$.
This cell and its image are shown in gray.
In this case it is possible to iterate $\sigma$ on its images;
a few iterations are shown below.
(The $\mcCs$-covering is drawn on each pattern.)
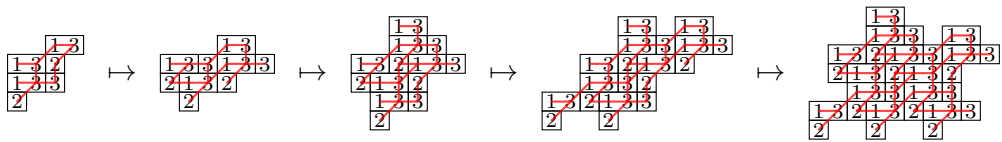
\[
\myvcenter{
%
}
\]
\end{exam}

\begin{exam}
\label{exam:inconsistent}
The substitution defined by the base rule
$1 \mapsto \{[(0,0),1\}$,
$2 \mapsto \{[(0,0),2\}$
and the concatenation rules
\myvcenter{%
}
is not consistent:
the $\mcCs$-loop
$\gamma = \{[(0,0), 2], [(1,0), 2], [(1,1), 1], [(0,1), 1], [(0,0), 2]\}$
has a nonzero image vector
$\ws(\gamma) = (1,-1) + (0,1) - (1,0) - (0,1) = (0,-1)$,
so $\sigma$ in inconsistent by Proposition~\ref{prop:loopchar}.
This is also illustrated by the following:
\begin{center}
\myvcenter{%
}\,.
\end{center}
\end{exam}

\begin{exam}
\label{exam:overlapping}
The substitution defined by the base rule
$1 \mapsto \{[(0,0),1\}$
and the concatenation rules
\myvcenter{%
}
is overlapping:
the images of the cells $[(0,1),1]$ and $[(1,0),1]$ overlap
in the image of the pattern $\{[(0,0),1], [(1,0),1], [(0,1),1]\}$:
\begin{center}
\myvcenter{%
}
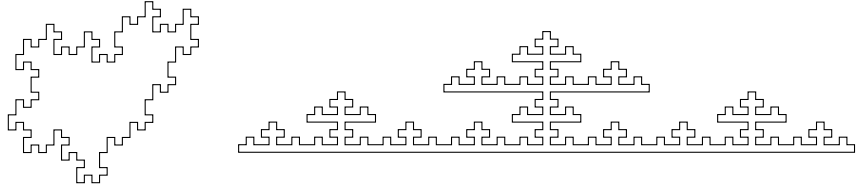
\caption{The boundaries of the patterns
$\sigma^{12}(\{[(0,0), 3], [(1,0), 1], [(1,1), 2]\}$
of Example~\ref{exam:JP} (left), and
$\sigma^4(\{[(0,0),1]\})$
of Example~\ref{exam:T} (right).
}
\end{figure}

\begin{exam}
The following example concerns the definition of consistency.
Let $\sigma$ be the substitution on the alphabet $\mcA = \{1,2,3,4,5,6\}$,
with the base rule $\sbase(a) = [(0,0),1]$ for every $a \in \mcA$,
and the concatenation rules
\[
\begin{array}{rcl}
(a,b,(1,0)) & \mapsto & (1,0) \ \text{ for } \ (a,b) \in \{(1,1), (2,2), (3,1), (4,2), (1,5), (2,6)\} \\
(3,4,(0,1)) & \mapsto & (0,1) \\
(5,6,(0,1)) & \mapsto & (0,-1).
\end{array}
\]
No $\mcCs$-covered-pattern can have more than two rows.
Also, the only patterns that admit more than one possible $\mcCs$-covering are the patterns of the form
$%
$,
but all their possible images are equal up to translation.
Hence, every $\mcCs$-covered pattern admits only one image by $\sigma$ up to translation.

This naturally suggests that $\sigma$ is consistent,
but it is in fact not the case: the two paths going
from the cell of type $4$ to the cell of type $5$ in the above pattern
do not have the same image vector.

Therefore, the property ``every $\mcCs$-covered pattern admits only one image up to translation''
is not equivalent to the consistency of $\sigma$,
which is an important fact to mention because it is
another natural candidate to define consistency.
Note that our definition of consistency is actually stronger.
Also, the undecidability proofs of Section~\ref{sec:und}
can easily be adapted to yield the undecidability of the alternative notion of consistency.
\end{exam}

\section{Undecidability results}
\label{sec:und}
\emph{Wang tiles} are unit square tiles with colored edges,
which are oriented and may not be rotated.
We say that a set of Wang tiles $T$ admits a valid tiling of a cycle if
there exists a nontrivial sequence $(a_1, \ldots, a_n)$ of translates of tiles of $T$
such that $a_i$ and $a_{i+1}$ share exactly one edge and their colors agree on it for all $1 \leq i < n$,
and such that $a_n = a_1$ (the other tiles $a_i$ cannot overlap and are distinct).
Because we require the cycle to be nontrivial, we must have $n \geq 5$.

In \cite{Kar02}, it is proved that the following problem is undecidable:
``Does a given finite set of Wang tiles admit a valid tiling of a cycle?''
This problem is called the \emph{weak cycle tiling problem} in \cite{Kar02}.
(The strong version of the same problem requires that \emph{any} two adjacent tiles in the cycle match in color,
and not only $a_i$ and $a_{i+1}$.)
We will use the fact that this problem is undecidable
in order to prove Theorems~\ref{theo:undcons} and~\ref{theo:undnono}.

The undecidability results below are proved for two-dimensional substitutions.
The proofs can easily be modified to get undecidability in higher dimensions,
but dimension $1$ has to be ruled out.

\subsection{Undecidability of consistency}

\begin{theo}
\label{theo:undcons}
It is undecidable whether a substitution is consistent.
\end{theo}

\begin{proof}
We are going to reduce the cycle tiling problem for Wang tiles
to the consistency problem for substitutions.
Since the former is undecidable, the result will follow.

Let $T$ be a set of Wang tiles.
Let $\mcA = T \times \{\rightarrow, \uparrow, \leftarrow, \downarrow\}$,
and $\sigma$ be the substitution over alphabet $\mcA$ defined by
$\sbase(t) = [(0,0), t]$ for all $t \in \mcA$,
with the rules

\begin{center}
\begin{tikzpicture}[x=4.5mm, y=4.5mm]
\tileU{0}{3}{$a$}
\tileR{0}{4}{$b$}
\node at (2,3.9) {$\mapsto$};
\tileU{3}{3.5}{$a$}
\tileR{4}{3.5}{$b$}

\tileU{7}{3}{$a$}
\tileU{7}{4}{$b$}
\node at (9,3.9) {$\mapsto$};
\tileU{10}{3.5}{$a$}
\tileU{11}{3.5}{$b$}

\tileU{14}{3}{$a$}
\tileL{14}{4}{$b$}
\node at (16,3.9) {$\mapsto$};
\tileU{17}{3.5}{$a$}
\tileL{18}{3.5}{$b$}

\tileR{0}{0}{$a$}
\tileD{0}{1}{$b$}
\node at (2,0.9) {$\mapsto$};
\tileD{3}{0.5}{$b$}
\tileR{4}{0.5}{$a$}

\tileD{7}{0}{$a$}
\tileD{7}{1}{$b$}
\node at (9,0.9) {$\mapsto$};
\tileD{10}{0.5}{$b$}
\tileD{11}{0.5}{$a$}

\tileL{14}{0}{$a$}
\tileD{14}{1}{$b$}
\node at (16,0.9) {$\mapsto$};
\tileD{17}{0.5}{$b$}
\tileL{18}{0.5}{$a$}
\end{tikzpicture}
\end{center}
for every tiles $a, b \in T$ such that the edges match in
\myvcenter{%
\begin{tikzpicture}[x=3mm, y=3mm]
\tileB{0}{0}{$a$}
\tileB{0}{1}{$b$}
\end{tikzpicture}%
}\,,
and the rules
\begin{center}
\begin{tikzpicture}[x=4.5mm, y=4.5mm]
\tileR{0}{2}{$a$}
\tileD{1}{2}{$b$}
\node at (3,2.45) {$\mapsto$};
\tileR{4}{2}{$a$}
\tileD{5}{2}{$b$}

\tileR{8}{2}{$a$}
\tileR{9}{2}{$b$}
\node at (11,2.45) {$\mapsto$};
\tileR{12}{2}{$a$}
\tileR{13}{2}{$b$}

\tileR{16}{2}{$a$}
\tileU{17}{2}{$b$}
\node at (19,2.45) {$\mapsto$};
\tileR{20}{2}{$a$}
\tileU{21}{2}{$b$}

\tileU{0}{0}{$a$}
\tileL{1}{0}{$b$}
\node at (3,0.45) {$\mapsto$};
\tileL{4}{0}{$b$}
\tileU{5}{0}{$a$}

\tileL{8}{0}{$a$}
\tileL{9}{0}{$b$}
\node at (11,0.45) {$\mapsto$};
\tileL{12}{0}{$b$}
\tileL{13}{0}{$a$}

\tileD{16}{0}{$a$}
\tileL{17}{0}{$b$}
\node at (19,0.45) {$\mapsto$};
\tileL{20}{0}{$b$}
\tileD{21}{0}{$a$}
\end{tikzpicture}
\end{center}
for every tiles $a, b \in T$ such that the edges match in
\myvcenter{%
\begin{tikzpicture}[x=3mm, y=3mm]
\tileB{0}{0}{$a$}
\tileB{1}{0}{$b$}
\end{tikzpicture}%
}\,.
By definition of the above rules,
the set $\mcCs$ consists of all the valid dominoes of tiles of $T$,
where in each domino, exactly one of the two tiles points at the other.
The image of a domino by $\sbase$ is then the concatenation
of the pointing tile and the pointed tile, from left to right respectively,
as illustrated below in an image by $\sigma$ of a $\sigma$-loop of length $8$:

\begin{center}
\myvcenter{%
\begin{tikzpicture}[x=5mm, y=5mm]
\tileR{0}{2}{$a_1$}
\tileR{1}{2}{$a_2$}
\tileD{2}{2}{$a_3$}
\tileD{2}{1}{$a_4$}
\tileL{2}{0}{$a_5$}
\tileL{1}{0}{$a_6$}
\tileU{0}{0}{$a_7$}
\tileU{0}{1}{$a_8$}
\end{tikzpicture}%
}
\ $\longmapsto$ \
\myvcenter{%
\begin{tikzpicture}[x=5mm, y=5mm]
\tileR{0}{0}{$a_1$}
\tileR{1}{0}{$a_2$}
\tileD{2}{0}{$a_3$}
\tileD{3}{0}{$a_4$}
\tileL{4}{0}{$a_5$}
\tileL{5}{0}{$a_6$}
\tileU{6}{0}{$a_7$}
\tileU{7}{0}{$a_8$}
\end{tikzpicture}%
}\,.
\end{center}

We can now finish the proof by showing that $T$ admits a valid tiling of a cycle
if and only if $\sigma$ is not consistent.
Indeed, suppose that $T$ admits a valid tiling of a cycle $(a_1, \ldots, a_n)$.
To this cycle corresponds a $\mcCs$-loop $\gamma = (c_1, \ldots, c_n)$
where the type of each $c_i$ is $(a_i, d_i)$ and
the arrow $d_i$ points at the cell $c_{i+1}$, for $1 \leq i < n$
(and $d_n$ points at $c_1$).
However, we have $\ws(\gamma) = (n-1,0) \neq (0,0)$, so $\sigma$ is not consistent,
by Proposition~\ref{prop:loopchar}.

Conversely, if $T$ does not admit a valid tiling of a cycle
then there cannot exist any simple $\mcCs$-loop,
so $\sigma$ is consistent thanks to Proposition~\ref{prop:loopchar}.
\end{proof}

\begin{rema}
The above proof yields a stronger version of Theorem~\ref{theo:undcons}:
consistency is undecidable for two-dimensional domino-to-domino substitutions.

We can also prove that undecidability of consistency holds for non-overlapping substitutions,
by modifying the above reduction slightly.
Let us first note that the substitution produced in the above reduction
is not necessarily non-overlapping,
as can be seen for example if two cells point at a same arrow,
in which case the two pointing cells will overlap in an image by $\sigma$:
\begin{center}
\myvcenter{%
\begin{tikzpicture}[x=5mm, y=5mm]
\draw[-stealth, draw=black] (0,0.5) -- (1,0.5);
\draw[draw=black] (0,0) -- (0,1) -- (1,1) -- (1,0) -- cycle;
\node[] at (0.5,0.25) {\scriptsize $a$};
\draw[-stealth, draw=black] (1.5,2) -- (1.5,1);
\draw[draw=black] (1,1) -- (1,2) -- (2,2) -- (2,1) -- cycle;
\node[] at (1.25,1.5) {\scriptsize $b$};
\draw[-stealth, draw=black] (1,0.5) -- (2,0.5);
\draw[draw=black] (1,0) -- (1,1) -- (2,1) -- (2,0) -- cycle;
\node[] at (1.5,0.25) {\scriptsize $c$};
\end{tikzpicture}%
}
\ $\longmapsto$ \
\myvcenter{%
\begin{tikzpicture}[x=5mm, y=5mm]
\draw[-stealth, draw=black, shift={(-1.25,-1.25)}] (1.5,2) -- (1.5,1);
\draw[draw=black, shift={(-1.25,-1.25)}] (1,1) -- (1,2) -- (2,2) -- (2,1) -- cycle;
\node[shift={(-1.25,-1.25)}] at (1.25,1.5) {\scriptsize $b$};
\fill[fill=white,draw=black] (0,0) -- (0,1) -- (1,1) -- (1,0) -- cycle;
\draw[-stealth, draw=black] (0,0.5) -- (1,0.5);
\node[] at (0.5,0.25) {\scriptsize $a$};
\draw[-stealth, draw=black] (1,0.5) -- (2,0.5);
\draw[draw=black] (1,0) -- (1,1) -- (2,1) -- (2,0) -- cycle;
\node[] at (1.5,0.25) {\scriptsize $c$};
\end{tikzpicture}%
}\,.
\end{center}
Hence, we want to make sure that the image of a pattern $P$ can be computed
only if a cell of $P$ is pointed by \emph{at most one} other cell of $P$.

The new reduction is then the following
given a tile set $T$
let $\sigma$ be the two-dimensional substitution defined on the alphabet
\[
\mcA \ = \ T \times \{
\myvcenter{\begin{tikzpicture}[x=5mm, y=5mm]
\draw[-stealth, draw=black] (0,0.5) -- (1,0.5);
\end{tikzpicture}}\,,\,
\myvcenter{\begin{tikzpicture}[x=5mm, y=5mm]
\draw[-stealth, draw=black] (0.5,0) -- (0.5,1);
\end{tikzpicture}}\,,\,
\myvcenter{\begin{tikzpicture}[x=5mm, y=5mm]
\draw[stealth-, draw=black] (0,0.5) -- (1,0.5);
\end{tikzpicture}}\,,\,
\myvcenter{\begin{tikzpicture}[x=5mm, y=5mm]
\draw[stealth-, draw=black] (0.5,0) -- (0.5,1);
\end{tikzpicture}}\,,\,
\myvcenter{\begin{tikzpicture}[x=5mm, y=5mm]
\draw[-stealth, draw=black] (0.5,1) -- (0.5,0.5) -- (1,0.5);
\end{tikzpicture}}\,,\,
\myvcenter{\begin{tikzpicture}[x=5mm, y=5mm]
\draw[stealth-, draw=black] (0.5,1) -- (0.5,0.5) -- (1,0.5);
\end{tikzpicture}}\,,\,
\myvcenter{\begin{tikzpicture}[x=5mm, y=5mm]
\draw[-stealth, draw=black] (0.5,1) -- (0.5,0.5) -- (0,0.5);
\end{tikzpicture}}\,,\,
\myvcenter{\begin{tikzpicture}[x=5mm, y=5mm]
\draw[stealth-, draw=black] (0.5,1) -- (0.5,0.5) -- (0,0.5);
\end{tikzpicture}}\,,\,
\myvcenter{\begin{tikzpicture}[x=5mm, y=5mm]
\draw[-stealth, draw=black] (0.5,0) -- (0.5,0.5) -- (1,0.5);
\end{tikzpicture}}\,,\,
\myvcenter{\begin{tikzpicture}[x=5mm, y=5mm]
\draw[stealth-, draw=black] (0.5,0) -- (0.5,0.5) -- (1,0.5);
\end{tikzpicture}}\,,\,
\myvcenter{\begin{tikzpicture}[x=5mm, y=5mm]
\draw[-stealth, draw=black] (0.5,0) -- (0.5,0.5) -- (0,0.5);
\end{tikzpicture}}\,,\,
\myvcenter{\begin{tikzpicture}[x=5mm, y=5mm]
\draw[stealth-, draw=black] (0.5,0) -- (0.5,0.5) -- (0,0.5);
\end{tikzpicture}}\,.
\}
\]
and such that $\mcCs$ consist of all the valid dominoes of tiles of $T$
in which exactly one tile points at the other,
but a tile is allowed to point at another if and only if
the tip of the arrow of the pointing tile matches with the tail of
the arrow of the pointed tile.
The rules of $\sigma$ behave similarly as in the above reduction:
the pointed tile is put at the right of the pointing tile,
as shown in the following three examples:
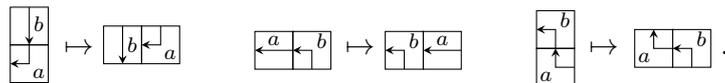
\begin{center}
\myvcenter{\begin{tikzpicture}[x=5mm, y=5mm]
\draw[draw=black] (0,0) -- (0,1) -- (1,1) -- (1,0) -- cycle;
\draw[-stealth, draw=black] (0.5,1) -- (0.5,0.5) -- (0,0.5);
\node[] at (0.75,0.25) {\scriptsize $a$};
\draw[draw=black] (0,1) -- (0,2) -- (1,2) -- (1,1) -- cycle;
\draw[-stealth, draw=black] (0.5,2) -- (0.5,1);
\node[] at (0.75,1.55) {\scriptsize $b$};
\end{tikzpicture}}
 $\mapsto$
\myvcenter{\begin{tikzpicture}[x=5mm, y=5mm]
\draw[draw=black] (1,0) -- (1,1) -- (2,1) -- (2,0) -- cycle;
\draw[-stealth, draw=black] (1.5,1) -- (1.5,0.5) -- (1,0.5);
\node[] at (1.75,0.25) {\scriptsize $a$};
\draw[draw=black] (0,0) -- (0,1) -- (1,1) -- (1,0) -- cycle;
\draw[-stealth, draw=black] (0.5,1) -- (0.5,0);
\node[] at (0.75,0.5) {\scriptsize $b$};
\end{tikzpicture}}
\qquad
\myvcenter{\begin{tikzpicture}[x=5mm, y=5mm]
\draw[draw=black] (0,0) -- (0,1) -- (1,1) -- (1,0) -- cycle;
\draw[stealth-, draw=black] (0,0.5) -- (1,0.5);
\node[] at (0.5,0.75) {\scriptsize $a$};
\draw[draw=black] (1,0) -- (1,1) -- (2,1) -- (2,0) -- cycle;
\draw[-stealth, draw=black] (1.5,0) -- (1.5,0.5) -- (1,0.5);
\node[] at (1.75,0.75) {\scriptsize $b$};
\end{tikzpicture}}
 $\mapsto$
\myvcenter{\begin{tikzpicture}[x=5mm, y=5mm]
\draw[draw=black] (1,0) -- (1,1) -- (2,1) -- (2,0) -- cycle;
\draw[stealth-, draw=black] (1,0.5) -- (2,0.5);
\node[] at (1.5,0.75) {\scriptsize $a$};
\draw[draw=black] (0,0) -- (0,1) -- (1,1) -- (1,0) -- cycle;
\draw[-stealth, draw=black] (0.5,0) -- (0.5,0.5) -- (0,0.5);
\node[] at (0.75,0.75) {\scriptsize $b$};
\end{tikzpicture}}
\qquad
\myvcenter{\begin{tikzpicture}[x=5mm, y=5mm]
\draw[draw=black] (0,0) -- (0,1) -- (1,1) -- (1,0) -- cycle;
\draw[stealth-, draw=black] (0.5,1) -- (0.5,0.5) -- (1,0.5);
\node[] at (0.25,0.25) {\scriptsize $a$};
\draw[draw=black] (0,1) -- (0,2) -- (1,2) -- (1,1) -- cycle;
\draw[-stealth, draw=black] (0.5,1) -- (0.5,1.5) -- (0,1.5);
\node[] at (0.75,1.75) {\scriptsize $b$};
\end{tikzpicture}}
 $\mapsto$
\myvcenter{\begin{tikzpicture}[x=5mm, y=5mm]
\draw[draw=black] (0,0) -- (0,1) -- (1,1) -- (1,0) -- cycle;
\draw[stealth-, draw=black] (0.5,1) -- (0.5,0.5) -- (1,0.5);
\node[] at (0.25,0.25) {\scriptsize $a$};
\draw[draw=black] (1,0) -- (1,1) -- (2,1) -- (2,0) -- cycle;
\draw[-stealth, draw=black] (1.5,0) -- (1.5,0.5) -- (1,0.5);
\node[] at (1.75,0.75) {\scriptsize $b$};
\end{tikzpicture}}\,.
\end{center}
The substitution is non-overlapping,
because the only patterns that admit an image by $\sigma$ are paths or cycles
decorated by matching arrows,
whose images are necessarily made of non-overlapping cells.
\end{rema}

\subsection{Undecidability of overlapping}

The weak cycle tiling problem can also be used to prove
the undecidability of overlapping for consistent two-dimensional substitutions.

\begin{theo}
\label{theo:undnono}
It is undecidable whether a consistent substitution is overlapping.
\end{theo}

\begin{proof}
We will reduce the cycle tiling problem.
Let $T$ be a set of Wang tiles.
Given two tiles $a, b \in T$ whose colors match in
\myvcenter{\begin{tikzpicture}[x=3.5mm, y=3.5mm]
\tileB{0}{0}{$a$}
\tileB{1}{0}{$b$}
\end{tikzpicture}},
let $\sigma_{a,b}$ be the two-dimensional substitution defined
on the alphabet $\mcA = T \cup \{a_0, b_0\}$
where $a_0$ and $b_0$ are two new states,
with the base rule $t \mapsto \{[(0,0), t]\}$ for all $t \in \mcA$
and the concatenation rules
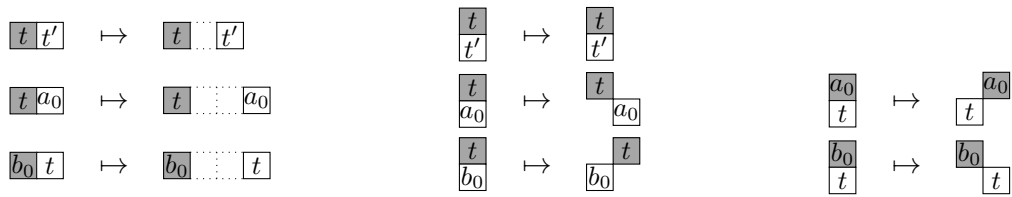
\begin{center}
$\begin{array}{rclcrclcrcl}
\myvcenter{\begin{tikzpicture}[x=3.5mm, y=3.5mm]
\tileBcol{0}{0}{$t$}{35}
\tileBcol{1}{0}{$t'$}{0}
\node at (1.5,0.5) {\small$\phantom{a_0}$}; 
\end{tikzpicture}}
& \mapsto &
\myvcenter{\begin{tikzpicture}[x=3.5mm, y=3.5mm]
\draw[draw=black, dotted] (1,0) -- (2,0);
\draw[draw=black, dotted] (1,1) -- (2,1);
\tileBcol{0}{0}{$t$}{35}
\tileBcol{2}{0}{$t'$}{0}
\node at (0.5,0.5) {\small$\phantom{b_0}$}; 
\end{tikzpicture}}
& \qquad \qquad &
\myvcenter{\begin{tikzpicture}[x=3.5mm, y=3.5mm]
\tileBcol{0}{1}{$t$}{35}
\tileBcol{0}{0}{$t'$}{0}
\node at (0.5,0.5) {\small$\phantom{a_0}$}; 
\end{tikzpicture}}
& \mapsto &
\myvcenter{\begin{tikzpicture}[x=3.5mm, y=3.5mm]
\tileBcol{0}{1}{$t$}{35}
\tileBcol{0}{0}{$t'$}{0}
\node at (0.5,0.5) {\small$\phantom{b_0}$}; 
\end{tikzpicture}}
& \qquad \qquad &
& & \\
\myvcenter{\begin{tikzpicture}[x=3.5mm, y=3.5mm]
\tileBcol{0}{0}{$t$}{35}
\tileBcol{1}{0}{$a_0$}{0}
\end{tikzpicture}}
& \mapsto &
\myvcenter{\begin{tikzpicture}[x=3.5mm, y=3.5mm]
\draw[draw=black, dotted] (1,0) -- (3,0);
\draw[draw=black, dotted] (1,1) -- (3,1);
\draw[draw=black, dotted] (2,0) -- (2,1);
\tileBcol{0}{0}{$t$}{35}
\tileBcol{3}{0}{$a_0$}{0}
\node at (0.5,0.5) {\small$\phantom{b_0}$}; 
\end{tikzpicture}}
& \qquad \qquad &
\myvcenter{\begin{tikzpicture}[x=3.5mm, y=3.5mm]
\tileBcol{0}{1}{$t$}{35}
\tileBcol{0}{0}{$a_0$}{0}
\end{tikzpicture}}
& \mapsto &
\myvcenter{\begin{tikzpicture}[x=3.5mm, y=3.5mm]
\tileBcol{0}{1}{$t$}{35}
\tileBcol{1}{0}{$a_0$}{0}
\node at (0.5,0.5) {\small$\phantom{b_0}$}; 
\end{tikzpicture}}
& \qquad \qquad &
\myvcenter{\begin{tikzpicture}[x=3.5mm, y=3.5mm]
\tileBcol{0}{1}{$a_0$}{35}
\tileBcol{0}{0}{$t$}{0}
\end{tikzpicture}}
& \mapsto &
\myvcenter{\begin{tikzpicture}[x=3.5mm, y=3.5mm]
\tileBcol{1}{1}{$a_0$}{35}
\tileBcol{0}{0}{$t$}{0}
\node at (0.5,0.5) {\small$\phantom{b_0}$}; 
\end{tikzpicture}} \\
\myvcenter{\begin{tikzpicture}[x=3.5mm, y=3.5mm]
\tileBcol{0}{0}{$b_0$}{35}
\tileBcol{1}{0}{$t$}{0}
\node at (1.5,0.5) {\small$\phantom{a_0}$}; 
\end{tikzpicture}}
& \mapsto &
\myvcenter{\begin{tikzpicture}[x=3.5mm, y=3.5mm]
\draw[draw=black, dotted] (1,0) -- (3,0);
\draw[draw=black, dotted] (1,1) -- (3,1);
\draw[draw=black, dotted] (2,0) -- (2,1);
\tileBcol{0}{0}{$b_0$}{35}
\tileBcol{3}{0}{$t$}{0}
\end{tikzpicture}}
& \qquad \qquad &
\myvcenter{\begin{tikzpicture}[x=3.5mm, y=3.5mm]
\tileBcol{0}{1}{$t$}{35}
\tileBcol{0}{0}{$b_0$}{0}
\node at (0.5,0.5) {\small$\phantom{a_0}$}; 
\end{tikzpicture}}
& \mapsto &
\myvcenter{\begin{tikzpicture}[x=3.5mm, y=3.5mm]
\tileBcol{1}{1}{$t$}{35}
\tileBcol{0}{0}{$b_0$}{0}
\end{tikzpicture}}
& \qquad \qquad &
\myvcenter{\begin{tikzpicture}[x=3.5mm, y=3.5mm]
\tileBcol{0}{1}{$b_0$}{35}
\tileBcol{0}{0}{$t$}{0}
\node at (0.5,0.5) {\small$\phantom{a_0}$}; 
\end{tikzpicture}}
& \mapsto &
\myvcenter{\begin{tikzpicture}[x=3.5mm, y=3.5mm]
\tileBcol{0}{1}{$b_0$}{35}
\tileBcol{1}{0}{$t$}{0}
\end{tikzpicture}}
\end{array}$
\end{center}
for the $t, t' \in \mcA \setminus \{a_0, b_0\}$
such that the tiles match in left-hand sides of the above rules
(we require $a_0$ and $b_0$ to match in the same way as $a$ and $b$).
On patterns without $a_0$ or $b_0$,
the image of a pattern by $\sigma_{a,b}$ is a copy expanded horizontally by a factor of two
(leaving one horizontal gap between horizontal neighbors).
When $a_0$ or $b_0$ is in the pattern,
the action of $\sigma_{a,b}$ is the same,
and in addition it shifts every occurrence of $a_0$ to the right
and every occurrence of $b_0$ to the left,
as illustrated below.
Note that the images of $a_0$ and $b_0$ are shifted
but that the images of $a$ and $b$ are not (they are treated like the other cells).
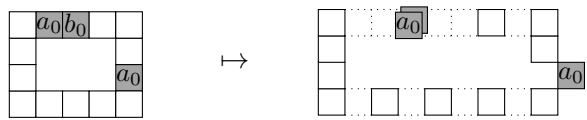
\begin{center}
\myvcenter{\begin{tikzpicture}[x=3.5mm, y=3.5mm]
\definecolor{cellcolor}{gray}{0.65}
\fill[fill=cellcolor, draw=black]
(1,3) -- (2,3) -- (2,4) -- (1,4) -- cycle;
\node at (1.5,3.5) {\small $a_0$};
\fill[fill=cellcolor, draw=black]
(2,3) -- (3,3) -- (3,4) -- (2,4) -- cycle;
\node at (2.5,3.5) {\small $b_0$};
\fill[fill=cellcolor, draw=black]
(4,1) -- (5,1) -- (5,2) -- (4,2) -- cycle;
\node at (4.5,1.5) {\small $a_0$};
\definecolor{cellcolor}{gray}{1}
\fill[fill=cellcolor, draw=black]
(0,0) -- (1,0) -- (1,1) -- (0,1) -- cycle;
\fill[fill=cellcolor, draw=black]
(1,0) -- (2,0) -- (2,1) -- (1,1) -- cycle;
\fill[fill=cellcolor, draw=black]
(2,0) -- (3,0) -- (3,1) -- (2,1) -- cycle;
\fill[fill=cellcolor, draw=black]
(3,0) -- (4,0) -- (4,1) -- (3,1) -- cycle;
\fill[fill=cellcolor, draw=black]
(4,0) -- (5,0) -- (5,1) -- (4,1) -- cycle;
\fill[fill=cellcolor, draw=black]
(0,1) -- (1,1) -- (1,2) -- (0,2) -- cycle;
\fill[fill=cellcolor, draw=black]
(0,2) -- (1,2) -- (1,3) -- (0,3) -- cycle;
\fill[fill=cellcolor, draw=black]
(4,2) -- (5,2) -- (5,3) -- (4,3) -- cycle;
\fill[fill=cellcolor, draw=black]
(0,3) -- (1,3) -- (1,4) -- (0,4) -- cycle;
\fill[fill=cellcolor, draw=black]
(3,3) -- (4,3) -- (4,4) -- (3,4) -- cycle;
\fill[fill=cellcolor, draw=black]
(4,3) -- (5,3) -- (5,4) -- (4,4) -- cycle;
\end{tikzpicture}}
$\qquad \mapsto \qquad$
\myvcenter{\begin{tikzpicture}[x=3.5mm, y=3.5mm]
\definecolor{cellcolor}{gray}{0.65}
\draw[draw=black, dotted] (1,0) -- (2,0);
\draw[draw=black, dotted] (1,1) -- (2,1);
\draw[draw=black, dotted] (3,0) -- (4,0);
\draw[draw=black, dotted] (3,1) -- (4,1);
\draw[draw=black, dotted] (5,0) -- (6,0);
\draw[draw=black, dotted] (5,1) -- (6,1);
\draw[draw=black, dotted] (7,0) -- (8,0);
\draw[draw=black, dotted] (7,1) -- (8,1);
\draw[draw=black, dotted] (7,3) -- (8,3);
\draw[draw=black, dotted] (7,4) -- (8,4);
\draw[draw=black, dotted] (1,3) -- (3,3);
\draw[draw=black, dotted] (1,4) -- (3,4);
\draw[draw=black, dotted] (2,3) -- (2,4);
\draw[draw=black, dotted] (4,3) -- (6,3);
\draw[draw=black, dotted] (4,4) -- (6,4);
\draw[draw=black, dotted] (5,3) -- (5,4);
\fill[fill=cellcolor, draw=black, shift={(.1,.1)}]
(3,3) -- (4,3) -- (4,4) -- (3,4) -- cycle;
\fill[fill=cellcolor, draw=black, shift={(-.1,-.1)}]
(3,3) -- (4,3) -- (4,4) -- (3,4) -- cycle;
\fill[fill=cellcolor, draw=black]
(9,1) -- (10,1) -- (10,2) -- (9,2) -- cycle;
\definecolor{cellcolor}{gray}{1}
\fill[fill=cellcolor, draw=black]
(0,0) -- (1,0) -- (1,1) -- (0,1) -- cycle;
\fill[fill=cellcolor, draw=black]
(2,0) -- (3,0) -- (3,1) -- (2,1) -- cycle;
\fill[fill=cellcolor, draw=black]
(4,0) -- (5,0) -- (5,1) -- (4,1) -- cycle;
\fill[fill=cellcolor, draw=black]
(6,0) -- (7,0) -- (7,1) -- (6,1) -- cycle;
\fill[fill=cellcolor, draw=black]
(8,0) -- (9,0) -- (9,1) -- (8,1) -- cycle;
\fill[fill=cellcolor, draw=black]
(0,1) -- (1,1) -- (1,2) -- (0,2) -- cycle;
\fill[fill=cellcolor, draw=black]
(0,2) -- (1,2) -- (1,3) -- (0,3) -- cycle;
\fill[fill=cellcolor, draw=black]
(8,2) -- (9,2) -- (9,3) -- (8,3) -- cycle;
\fill[fill=cellcolor, draw=black]
(0,3) -- (1,3) -- (1,4) -- (0,4) -- cycle;
\fill[fill=cellcolor, draw=black]
(6,3) -- (7,3) -- (7,4) -- (6,4) -- cycle;
\fill[fill=cellcolor, draw=black]
(8,3) -- (9,3) -- (9,4) -- (8,4) -- cycle;
\node at (9.5,1.5) {\small $a_0$};
\node at (3.4,3.4) {\small $a_0$};
\end{tikzpicture}}
\end{center}

The rule is hence consistent,
and an overlap can happen only between the images of
a cell of type $a_0$ and a cell of type $b_0$:
an overlap occurs if and only if the image of
\myvcenter{\begin{tikzpicture}[x=3.5mm, y=3.5mm]
\tileB{0}{0}{$a_0$}
\tileB{1}{0}{$b_0$}
\end{tikzpicture}}
is computed (as shown in the above picture).
It follows that $\sigma_{a,b}$ is overlapping if and only if there
exists a Wang tile cycle of $T$ that contains
\myvcenter{\begin{tikzpicture}[x=3.5mm, y=3.5mm]
\tileB{0}{0}{$a$}
\tileB{1}{0}{$b$}
\end{tikzpicture}}.
Indeed, if there exists such a cycle,
then the image of the corresponding pattern in which exactly one occurrence of
\myvcenter{\begin{tikzpicture}[x=3.5mm, y=3.5mm]
\tileB{0}{0}{$a$}
\tileB{1}{0}{$b$}
\end{tikzpicture}}
is replaced by
\myvcenter{\begin{tikzpicture}[x=3.5mm, y=3.5mm]
\tileB{0}{0}{$a_0$}
\tileB{1}{0}{$b_0$}
\end{tikzpicture}}
can be computed (is $\mcC_{\sigma_{a,b}}$-covered) and will cause an overlap.
Conversely,
if an overlap exists then we know that
it is because the image of
\myvcenter{\begin{tikzpicture}[x=3.5mm, y=3.5mm]
\tileB{0}{0}{$a_0$}
\tileB{1}{0}{$b_0$}
\end{tikzpicture}}
has been computed.
This is possible only if a cycle of $T$ containing
\myvcenter{\begin{tikzpicture}[x=3.5mm, y=3.5mm]
\tileB{0}{0}{$a$}
\tileB{1}{0}{$b$}
\end{tikzpicture}}
exists, because
\myvcenter{\begin{tikzpicture}[x=3.5mm, y=3.5mm]
\tileB{0}{0}{$a_0$}
\tileB{1}{0}{$b_0$}
\end{tikzpicture}}
is not a starting pattern of $\sigma_{a,b}$.

Now we can finish the reduction.
Given a set of Wang tiles $T$,
compute $\sigma_{a,b}$ for every tiles $a, b$ whose colors match in
\myvcenter{\begin{tikzpicture}[x=3.5mm, y=3.5mm]
\tileB{0}{0}{$a$}
\tileB{1}{0}{$b$}
\end{tikzpicture}}.
One of the substitutions $\sigma_{a,b}$ is overlapping
if and only if $T$ admits a tiling of a cycle.
\end{proof}

\begin{rema}
The above proof yields a stronger version of Theorem~\ref{theo:undnono}:
non-overlapping is undecidable for consistent two-dimensional domino substitutions.
One can also prove that this holds even for domino-to-domino substitutions.
\end{rema}

\section{Decidability results}
\label{sect:dec}
In this section we give algorithms to decide the consistency or the non-overlapping
of a natural class of substitutions:
the substitutions $\sigma$ that are \emph{domino-complete},
that is, such that the set of starting patterns $\mcCs$
is the set of \emph{all} the possible dominoes.

\subsection{Decidability of consistency for domino-complete substitutions}

\begin{theo}
\label{theo:deccons}
It is decidable whether a given
two-dimensional domino-complete substitution is consistent.
More precisely, such a substitution is consistent if and only if
it is consistent on every $2 \times 2$ pattern.
\end{theo}

\begin{proof}
The ``only if'' implication is trivial.
For the ``if'' implication, suppose that $\sigma$ is not consistent.
By Proposition~\ref{prop:loopchar},
there exists a simple $\mcCs$-loop $\gamma = (c_1, \ldots, c_n)$ such that $\ws(\gamma) \neq 0$.
We will prove that there exists a $2 \times 2$ pattern on which $\sigma$ is not consistent
by ``reducing'' $\gamma$ inductively.

Let $c$ be the lowest cell on the leftmost column of $\gamma$.
Since $\gamma$ does not overlap itself,
there exist two cells $d, e \in \gamma$ such that
$d$ is above $c$ and $e$ is at the right of $c$.
We suppose, without loss of generality, that $d, c, e$ appear in this order in $\gamma$, \emph{i.e.},
$\gamma = (c_1, \ldots, c_i, d, c, e, c_{i+4}, \ldots, c_n)$.
Let $f = [v + (1,1), t]$,
where $v$ is the vector of $c$ and $t \in \mcA$ is arbitrary
(or $t$ agrees with $\gamma$ if $\gamma$ already contains
a cell of vector $v + (1,1)$).
Let $\gamma' = (c, e, f, d, c)$
and $\gamma'' = (c_1, \ldots, c_i, d, f, e, c_{i+4}, \ldots, c_n)$,
as shown below.
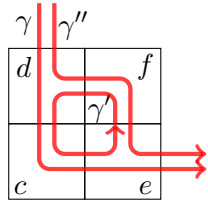
\begin{center}
\begin{tikzpicture}
\tileB{0}{1}{}
\tileB{0}{0}{}
\tileB{1}{0}{}
\tileB{1}{1}{}
\node at (0.15,0.15) {$c$};
\node at (0.2,1.75) {$d$};
\node at (1.8,0.15) {$e$};
\node at (1.8,1.75) {$f$};
\node at (0.2,2.30) {$\gamma$};
\node at (0.85,2.30) {$\gamma''$};
\node at (1.2,1.2) {$\gamma'$};
\draw[->, ultra thick, draw=red, rounded corners, opacity=0.75]
    (1.4,1) -- (1.4,1.4) -- (0.6,1.4) -- (0.6,0.6) -- (1.4,0.6) -- (1.4,1);
\draw[->, ultra thick, draw=red, rounded corners, opacity=0.75]
    (0.6,2.6) -- (0.6,1.6) -- (1.6,1.6) -- (1.6,0.6) -- (2.6,0.6);
\draw[->, ultra thick, draw=red, rounded corners, opacity=0.75]
    (0.4,2.6) -- (0.4,1.6) -- (0.4,0.4) -- (1.6,0.4) -- (2.6,0.4);
\end{tikzpicture}
\end{center}
We have $\ws(\gamma') + \ws(\gamma'') = \ws(\gamma) \neq 0$,
so $\ws(\gamma') \neq 0$ or $\ws(\gamma'') \neq 0$,
which implies the existence of a $\mcCs$-loop ($\gamma'$ or $\gamma''$)
with nonzero image vector which surrounds strictly less cells than $\gamma$
(unless $\gamma$ consists $4$ cells already).
Now, in the same way as in the second part of the proof of Proposition~\ref{prop:loopchar},
$\gamma'$ or $\gamma''$ must contain a simple loop with nonzero image vector.
Applying this reasoning inductively eventually leads to
a $2 \times 2$ loop $\gamma$ such that $\ws(\gamma) \neq 0$,
which concludes the proof.
\end{proof}

\paragraph{Generalization to domino-completeness within a set of patterns}
We now want to generalize Theorem~\ref{theo:deccons} to substitutions
that are domino-complete only within a particular set of patterns.
Let us first state a few definitions.
If $\mcP$ is a set of patterns,
we say that $\sigma$ is \emph{$\mcP$-domino-complete}
if $\mcCs$ is equal to the set of dominoes that appear in the patterns of $\mcP$.
If $\mcS \subseteq \mcA^{\bbZ^2}$,
we denote by $\patt(\mcS)$ the set of the patterns that appear in the elements of $\mcS$.
(That is, a pattern is in $\patt(\mcS)$ if it can be extended to an element of $\mcS$.)

Theorem~\ref{theo:decconsmieux} below gives a simple criterion
to determine if a $\mcP$-domino-complete substitution is consistent
when $\mcP$ is the set of all the $2 \times 2$ patterns of some $\mcS \subseteq \mcA^{\bbZ^2}$.
Note that to decide this property, we must be able to compute the $2 \times 2$
patterns of $\mcS$, which is not necessarily possible.
Note that Theorem~\ref{theo:deccons} can be seen as the particular case of Theorem~\ref{theo:decconsmieux}
when $\mcS = \mcA^{\bbZ^2}$.

\begin{theo}
\label{theo:decconsmieux}
Let $\mcS \subseteq \mcA^{\bbZ^2}$
and let $\mcP = \{P_1, \ldots, P_n\}$
be the list of the $2 \times 2$ patterns that appear in the elements of $\mcS$.
A $\mcP$-domino-complete substitution is consistent on $\patt(\mcS)$
if and only if it is consistent on the patterns $P_1, \ldots, P_n$.
\end{theo}

\begin{proof}
Let $P$ be a pattern of $\patt(\mcS)$ that contains a $\mcCs$-loop $\gamma$ such that $\ws(\gamma) \neq 0$.
We cannot directly reduce the loop as in the proof of Theorem~\ref{theo:deccons}
because $\sigma$ is not domino-complete.
However, $\sigma$ is $\mcP$-domino complete
so there exists $c \in \mcS$ that contains $P$.
We can then reduce $\gamma$ within $c$ to a $2 \times 2$ loop,
as explained in the proof of Theorem~\ref{theo:deccons}.
It follows that $\sigma$ is consistent on $\patt(\mcS)$
if and only if it is consistent on $P_1, \ldots, P_n$.
\end{proof}

\subsection{Decidability of overlapping for consistent domino-complete substitutions}
\label{sect:decoverlap}
We now focus on the domino-complete substitutions that are consistent.
Proposition~\ref{prop:structcdc} below tells us that such substitutions are simple:
there exists $\alpha, \beta \in \bbZ^2$ such that
the image of the cell placed at $(x,y)$ is placed at $(x\alpha, y\beta)$
in the lattice $\alpha\bbZ \times \beta\bbZ$.
We will use this proposition to give an algorithm that decides
if a consistent domino-complete is overlapping (Theorem~\ref{theo:decnono}).

Unfortunately, we are not able to give an analogue of Theorem~\ref{theo:decconsmieux}
for the non-overlapping property,
because the associated decision problems seem to become too difficult to track.

\begin{prop}
\label{prop:structcdc}
Let $\sigma$ be a consistent two-dimensional domino-complete substitution.
There exist two vectors $\alpha, \beta \in \bbZ^2$
and a vector $v_t \in \bbZ^2$ for every $t \in \mcA$
such that for every $\mcCs$-path $\gamma$
from a cell $[(0,0), t]$ to a cell $[(x,y), t']$,
we have $\ws(\gamma) = x \alpha + y \beta - v_t + v_{t'}$.
Moreover, $\alpha$, $\beta$ and $v_t$ can be obtained effectively.
\end{prop}

\begin{proof}
Let $t_0 \in \mcA$ be arbitrary, where $\mcA$ is the alphabet of $\sigma$.
Let $\alpha = \ws(\gamma)$ and $\beta = \ws(\gamma')$,
where $\gamma = ([(0,0), t_0], [(1,0), t_0])$ and
$\gamma' = ([(0,0), t_0], [(0,1), t_0])$.
For $t \in \mcA$, define $v_t = \ws(\gamma) - \alpha$,
where $\gamma = ([(0,0), t_0], [(1,0), t])$.
We first prove the theorem for paths of length two.
Because $\sigma$ is consistent and domino-complete, we can use the following patterns
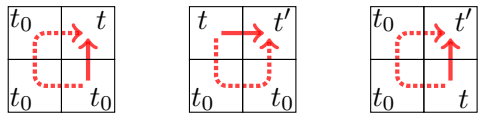
\[
\myvcenter{\begin{tikzpicture}[x=7mm, y=7mm]
\draw[->, ultra thick, draw=red, rounded corners, densely dotted, opacity=0.75]
    (1.4,0.5) -- (0.5,0.5) -- (0.5,1.5) -- (1.4,1.5);
\draw[->, ultra thick, draw=red, rounded corners, opacity=0.75]
    (1.5,0.6) -- (1.5,1.4);
\draw[black]
(0, 1) -- (0, 0) -- (1, 0) -- (1, 1) -- cycle;
\node at (0.25,0.25) {$t_0$};
\draw[black]
(1, 1) -- (1, 0) -- (2, 0) -- (2, 1) -- cycle;
\node at (1.75,0.25) {$t_0$};
\draw[black]
(0, 2) -- (0, 1) -- (1, 1) -- (1, 2) -- cycle;
\node at (0.25,1.75) {$t_0$};
\draw[black]
(1, 2) -- (1, 1) -- (2, 1) -- (2, 2) -- cycle;
\node at (1.75,1.75) {$t$};
\end{tikzpicture}}
\qquad
\myvcenter{\begin{tikzpicture}[x=7mm, y=7mm]
\draw[->, ultra thick, draw=red, rounded corners, densely dotted, opacity=0.75]
    (0.5,1.4) -- (0.5,0.5) -- (1.5,0.5) -- (1.5,1.4);
\draw[->, ultra thick, draw=red, rounded corners, opacity=0.75]
    (0.6,1.5) -- (1.4,1.5);
\draw[black]
(0, 1) -- (0, 0) -- (1, 0) -- (1, 1) -- cycle;
\node at (0.25,0.25) {$t_0$};
\draw[black]
(1, 1) -- (1, 0) -- (2, 0) -- (2, 1) -- cycle;
\node at (1.75,0.25) {$t_0$};
\draw[black]
(0, 2) -- (0, 1) -- (1, 1) -- (1, 2) -- cycle;
\node at (0.25,1.75) {$t$};
\draw[black]
(1, 2) -- (1, 1) -- (2, 1) -- (2, 2) -- cycle;
\node at (1.75,1.75) {$t'$};
\end{tikzpicture}}
\qquad
\myvcenter{\begin{tikzpicture}[x=7mm, y=7mm]
\draw[->, ultra thick, draw=red, rounded corners, densely dotted, opacity=0.75]
    (1.4,0.5) -- (0.5,0.5) -- (0.5,1.5) -- (1.4,1.5);
\draw[->, ultra thick, draw=red, rounded corners, opacity=0.75]
    (1.5,0.6) -- (1.5,1.4);
\draw[black]
(0, 1) -- (0, 0) -- (1, 0) -- (1, 1) -- cycle;
\node at (0.25,0.25) {$t_0$};
\draw[black]
(1, 1) -- (1, 0) -- (2, 0) -- (2, 1) -- cycle;
\node at (1.75,0.25) {$t$};
\draw[black]
(0, 2) -- (0, 1) -- (1, 1) -- (1, 2) -- cycle;
\node at (0.25,1.75) {$t_0$};
\draw[black]
(1, 2) -- (1, 1) -- (2, 1) -- (2, 2) -- cycle;
\node at (1.75,1.75) {$t'$};
\end{tikzpicture}}
\]
to compute the values
\[
\begin{array}{lclcl}
\ws([(0,0), t_0], [(0,1), t]) & = & -\alpha + \beta + \alpha + v_t & = & \beta + v_t \\
\ws([(0,0), t], [(1,0), t']) & = & -\beta - v_t + \alpha + \beta + v_{t'} & = & \alpha - v_t + v_{t'} \\
\ws([(0,0), t], [(0,1), t']) & = & -\alpha - v_t + \beta + \alpha + v_{t'} & = & \beta - v_t + v_{t'}
\end{array}
\]
which proves the statement for paths of length two.
The statement for arbitrary paths follows directly,
by adding the consecutive dominoes along the path.
\end{proof}

\begin{theo}
\label{theo:decnono}
It is decidable whether a given
two-dimensional consistent domino-complete substitution $\sigma$ is overlapping.
\end{theo}

\begin{proof}
By Proposition~\ref{prop:structcdc},
we can compute $\alpha, \beta, v_t, v_{t'} \in \bbZ^2$ such that
$\ws(\gamma) = x \alpha + y \beta - v_t + v_{t'}$
for every $\mcCs$-path $\gamma$
from a cell $[(0,0),t]$ to a cell $[(x,y),t']$.

Denote $A_t = \supp(\sbase(t))$ for $t \in \mcA$.
By definition, $\sigma$ is overlapping if and only if
there exists $t, t' \in \mcA$ and $x, y \in \bbZ$ such that
$A_t \cap \{b + x \alpha + y \beta - v_t + v_{t'} : b \in A_{t'}\} \neq \varnothing$.
This leaves a finite number of linear equations to check:
for each $(t, t') \in \mcA^2$, we check if there exists
$a \in A_t$ and $b \in A_{t'}$
such that the following equation has a nonzero solution $(x,y) \in \bbZ^2$:
\[
a \ = \ b + x \alpha + y \beta - v_t + v_{t'}
\]
This can be done algorithmically and $\sigma$ is overlapping if and only if such a solution exists.
\end{proof}

\begin{exam}
\label{exam:T}
Let $\sigma$ be the two-dimensional substitution on the alphabet $\{1\}$
defined by the base rule
\[
1 \ \mapsto \ \{[(0,0), 1], [(1,0), 1], [(2,0), 1], [(1,1), 1]\} \ = \
\myvcenter{\begin{tikzpicture}[x=3mm, y=3mm]
\draw[black]
(0, 1) -- (0, 0) -- (1, 0) -- (1, 1) -- cycle;
\node at (0.5,0.5) {\scriptsize $1$};
\draw[black]
(1, 1) -- (1, 0) -- (2, 0) -- (2, 1) -- cycle;
\node at (1.5,0.5) {\scriptsize $1$};
\draw[black]
(2, 1) -- (2, 0) -- (3, 0) -- (3, 1) -- cycle;
\node at (2.5,0.5) {\scriptsize $1$};
\draw[black]
(1, 2) -- (1, 1) -- (2, 1) -- (2, 2) -- cycle;
\node at (1.5,1.5) {\scriptsize $1$};
\end{tikzpicture}}
\]
and the concatenation rules
$(1, 1, (1,0)) \mapsto (3,0)$ and
$(1, 1, (0,1)) \mapsto (0,2)$.
This substitution is domino-complete and consistent.
Proposition~\ref{prop:structcdc} applied to $\sigma$
gives $\ws(\gamma) = (3x,0) + (0,2y)$ for every $\mcCs$-path $\gamma$
from a cell $[(0,0),1]$ to a cell $[(x,y),1]$.
Note that in this case, $v_1 = (0,0)$.
\end{exam}

\begin{exam}
\label{exam:overlapfar}
For $n \geq 0$, let $\sigma_n$ be the two-dimensional substitution on the alphabet $\{1,2\}$
defined by the base rule
$1 \mapsto \{[(0,0), 1]\}$,
$2 \mapsto \{[(0,0), 2]\}$
and
\begin{center}
\begin{tabular}{rclcrclcrcl}
\myvcenter{\begin{tikzpicture}[x=3mm, y=3mm]
\draw[black]
(0, 1) -- (0, 0) -- (1, 0) -- (1, 1) -- cycle;
\node at (0.5,0.5) {\scriptsize $a$};
\definecolor{cellcolor}{rgb}{0.65,0.65,0.65}
\fill[fill=cellcolor, draw=black]
(1, 1) -- (1, 0) -- (2, 0) -- (2, 1) -- cycle;
\node at (1.5,0.5) {\scriptsize $a$};
\end{tikzpicture}}
& $\mapsto$ &
\myvcenter{\begin{tikzpicture}[x=3mm, y=3mm]
\draw[black]
(0, 1) -- (0, 0) -- (1, 0) -- (1, 1) -- cycle;
\node at (0.5,0.5) {\scriptsize $a$};
\definecolor{cellcolor}{rgb}{0.65,0.65,0.65}
\fill[fill=cellcolor, draw=black]
(1, 1) -- (1, 0) -- (2, 0) -- (2, 1) -- cycle;
\node at (1.5,0.5) {\scriptsize $a$};
\end{tikzpicture}}
    & &
\myvcenter{\begin{tikzpicture}[x=3mm, y=3mm]
\draw[black]
(0, 1) -- (0, 0) -- (1, 0) -- (1, 1) -- cycle;
\node at (0.5,0.5) {\scriptsize $1$};
\definecolor{cellcolor}{rgb}{0.65,0.65,0.65}
\fill[fill=cellcolor, draw=black]
(1, 1) -- (1, 0) -- (2, 0) -- (2, 1) -- cycle;
\node at (1.5,0.5) {\scriptsize $2$};
\end{tikzpicture}}
& $\mapsto$ &
\myvcenter{\begin{tikzpicture}[x=3mm, y=3mm]
\draw[black]
(-4, 0) -- (-4, 1) -- (-5, 1) -- (-5, 0) -- cycle;
\node at (-4.5,0.5) {\scriptsize $1$};
\definecolor{cellcolor}{rgb}{0.65,0.65,0.65}
\fill[fill=cellcolor, draw=black]
(0, 1) -- (0, 0) -- (1, 0) -- (1, 1) -- cycle;
\node at (0.5,0.5) {\scriptsize $2$};
\draw[dashed, black, |<->|]
(-5,1.3) -- node [above] {\scriptsize $n+1$} (0,1.3);
\end{tikzpicture}}
    & &
\myvcenter{\begin{tikzpicture}[x=3mm, y=3mm]
\draw[black]
(0, 1) -- (0, 0) -- (1, 0) -- (1, 1) -- cycle;
\node at (0.5,0.5) {\scriptsize $1$};
\definecolor{cellcolor}{rgb}{0.65,0.65,0.65}
\fill[fill=cellcolor, draw=black]
(0, 2) -- (0, 1) -- (1, 1) -- (1, 2) -- cycle;
\node at (0.5,1.5) {\scriptsize $2$};
\end{tikzpicture}}
& $\mapsto$ &
\myvcenter{\begin{tikzpicture}[x=3mm, y=3mm]
\draw[black]
(-3, -1) -- (-3, 0) -- (-4, 0) -- (-4, -1) -- cycle;
\node at (-3.5,-0.5) {\scriptsize $1$};
\definecolor{cellcolor}{rgb}{0.65,0.65,0.65}
\fill[fill=cellcolor, draw=black]
(0, 1) -- (0, 0) -- (1, 0) -- (1, 1) -- cycle;
\node at (0.5,0.5) {\scriptsize $2$};
\draw[dashed, black, |<->|]
(-4,0.3) -- node [above] {\scriptsize $n$} (0,0.3);
\end{tikzpicture}}
    \\
\myvcenter{\begin{tikzpicture}[x=3mm, y=3mm]
\draw[black]
(0, 1) -- (0, 0) -- (1, 0) -- (1, 1) -- cycle;
\node at (0.5,0.5) {\scriptsize $a$};
\definecolor{cellcolor}{rgb}{0.65,0.65,0.65}
\fill[fill=cellcolor, draw=black]
(0, 2) -- (0, 1) -- (1, 1) -- (1, 2) -- cycle;
\node at (0.5,1.5) {\scriptsize $a$};
\end{tikzpicture}}
& $\mapsto$ &
\myvcenter{\begin{tikzpicture}[x=3mm, y=3mm]
\draw[black]
(0, 0) -- (0, -1) -- (1, -1) -- (1, 0) -- cycle;
\node at (0.5,-0.5) {\scriptsize $a$};
\definecolor{cellcolor}{rgb}{0.65,0.65,0.65}
\fill[fill=cellcolor, draw=black]
(0, 1) -- (0, 0) -- (1, 0) -- (1, 1) -- cycle;
\node at (0.5,0.5) {\scriptsize $a$};
\end{tikzpicture}}
    & &
\myvcenter{\begin{tikzpicture}[x=3mm, y=3mm]
\draw[black]
(0, 1) -- (0, 0) -- (1, 0) -- (1, 1) -- cycle;
\node at (0.5,0.5) {\scriptsize $2$};
\definecolor{cellcolor}{rgb}{0.65,0.65,0.65}
\fill[fill=cellcolor, draw=black]
(1, 1) -- (1, 0) -- (2, 0) -- (2, 1) -- cycle;
\node at (1.5,0.5) {\scriptsize $1$};
\end{tikzpicture}}
& $\mapsto$ &
\myvcenter{\begin{tikzpicture}[x=3mm, y=3mm]
\draw[black]
(3, 1) -- (3, 0) -- (4, 0) -- (4, 1) -- cycle;
\node at (3.5,0.5) {\scriptsize $2$};
\definecolor{cellcolor}{rgb}{0.65,0.65,0.65}
\fill[fill=cellcolor, draw=black]
(0, 1) -- (0, 0) -- (1, 0) -- (1, 1) -- cycle;
\node at (0.5,0.5) {\scriptsize $1$};
\draw[dashed, black, |<->|]
(0,1.3) -- node [above] {\scriptsize $n-1$} (3,1.3);
\end{tikzpicture}}
    & &
\myvcenter{\begin{tikzpicture}[x=3mm, y=3mm]
\draw[black]
(0, 1) -- (0, 0) -- (1, 0) -- (1, 1) -- cycle;
\node at (0.5,0.5) {\scriptsize $2$};
\definecolor{cellcolor}{rgb}{0.65,0.65,0.65}
\fill[fill=cellcolor, draw=black]
(0, 2) -- (0, 1) -- (1, 1) -- (1, 2) -- cycle;
\node at (0.5,1.5) {\scriptsize $1$};
\end{tikzpicture}}
& $\mapsto$ &
\myvcenter{\begin{tikzpicture}[x=3mm, y=3mm]
\draw[black]
(4, 0) -- (4, -1) -- (5, -1) -- (5, 0) -- cycle;
\node at (4.5,-0.5) {\scriptsize $2$};
\definecolor{cellcolor}{rgb}{0.65,0.65,0.65}
\fill[fill=cellcolor, draw=black]
(0, 1) -- (0, 0) -- (1, 0) -- (1, 1) -- cycle;
\node at (0.5,0.5) {\scriptsize $1$};
\draw[dashed, black, |<->|]
(0,-0.3) -- node [above] {\scriptsize $n$} (4,-0.3);
\end{tikzpicture}}
\end{tabular}
\end{center}
The substitution $\sigma_n$ is domino-complete and consistent for all $n$.
Proposition~\ref{prop:structcdc} applied to $\sigma_n$
gives $\ws(\gamma) = (x,y) - v_t + v_{t'}$ for every $\mcCs$-path $\gamma$
from a cell $[(0,0),t]$ to a cell $[(x,y),t']$,
where $v_1 = (0,0)$ and $v_2 = (n,0)$.
This gives an example of a substitution with at least one nonzero $v_t$.

This example is also interesting because it is overlapping, but only on sufficiently large patterns.
Indeed, it is non-overlapping on patterns of horizontal diameter smaller than $n$,
but overlapping on larger patterns such as
\[
P_n \ = \
\myvcenter{%
\begin{tikzpicture}[x=3mm, y=3mm]
\draw[black]
(1, 1) -- (1, 2) -- (2, 2) -- (2, 1) -- cycle;
\draw[black]
(2, 1) -- (2, 2) -- (3, 2) -- (3, 1) -- cycle;
\draw[black]
(3, 1) -- (3, 2) -- (4, 2) -- (4, 1) -- cycle;
\draw[black]
(6, 1) -- (6, 2) -- (7, 2) -- (7, 1) -- cycle;
\draw[black]
(7, 1) -- (7, 2) -- (8, 2) -- (8, 1) -- cycle;
\draw[black]
(7, 0) -- (7, 1) -- (8, 1) -- (8, 0) -- cycle;
\node at (1.5,1.5) {\scriptsize $1$};
\node at (2.5,1.5) {\scriptsize $1$};
\node at (3.5,1.5) {\scriptsize $1$};
\node at (5,1.5) {\scriptsize $\cdots$};
\node at (6.5,1.5) {\scriptsize $1$};
\node at (7.5,1.5) {\scriptsize $1$};
\node at (7.5,0.5) {\scriptsize $1$};
\definecolor{cellcolor}{rgb}{0.65,0.65,0.65}
\fill[fill=cellcolor, draw=black]
(1, 1) -- (1, 0) -- (2, 0) -- (2, 1) -- cycle;
\node at (1.5,0.5) {\scriptsize $2$};
\draw[dashed, black, |<->|]
(1,-0.3) -- node [below] {\scriptsize $n$} (7,-0.3);
\end{tikzpicture}}
\ = \ [(0,0),2] \cup [(n+1,0),1] \cup \{[(i,1),1] : 0 \leq i \leq n+1\}.
\]
This shows that the overlapping property cannot be decided as simply as consistency,
where looking at the $2 \times 2$ patterns was sufficient.
Now the size of the rules has to be taken into account,
which explains why the algorithm of Theorem~\ref{theo:decnono}
is not as simple as the algorithm of Theorem~\ref{theo:deccons}.
\end{exam}

\subsection{Applications}
\label{sec:applis}
Let $\Ssurf \subseteq \{1,2,3\}^{\bbZ^2}$ be the set of elements of $\{1,2,3\}^{\bbZ^2}$
whose set of allowed $2 \times 2$ patterns is the following set $\Psurf$ of $28$ patterns
\begin{center}
$
\myvcenter{%
\begin{tikzpicture}[x=2.5mm, y=2.5mm]
\draw[black]
(0, 1) -- (0, 0) -- (1, 0) -- (2, 0) -- (2, 1) -- (2, 2) -- (1, 2) -- (0, 2) -- cycle;
\node at (0.5,0.5) {\scriptsize $1$};
\node at (1.5,0.5) {\scriptsize $1$};
\node at (0.5,1.5) {\scriptsize $1$};
\node at (1.5,1.5) {\scriptsize $1$};
\end{tikzpicture} \
\begin{tikzpicture}[x=2.5mm, y=2.5mm]
\draw[black]
(0, 1) -- (0, 0) -- (1, 0) -- (2, 0) -- (2, 1) -- (2, 2) -- (1, 2) -- (0, 2) -- cycle;
\node at (0.5,0.5) {\scriptsize $1$};
\node at (1.5,0.5) {\scriptsize $1$};
\node at (0.5,1.5) {\scriptsize $2$};
\node at (1.5,1.5) {\scriptsize $1$};
\end{tikzpicture} \
\begin{tikzpicture}[x=2.5mm, y=2.5mm]
\draw[black]
(0, 1) -- (0, 0) -- (1, 0) -- (2, 0) -- (2, 1) -- (2, 2) -- (1, 2) -- (0, 2) -- cycle;
\node at (0.5,0.5) {\scriptsize $1$};
\node at (1.5,0.5) {\scriptsize $1$};
\node at (0.5,1.5) {\scriptsize $3$};
\node at (1.5,1.5) {\scriptsize $1$};
\end{tikzpicture} \
\begin{tikzpicture}[x=2.5mm, y=2.5mm]
\draw[black]
(0, 1) -- (0, 0) -- (1, 0) -- (2, 0) -- (2, 1) -- (2, 2) -- (1, 2) -- (0, 2) -- cycle;
\node at (0.5,0.5) {\scriptsize $1$};
\node at (1.5,0.5) {\scriptsize $2$};
\node at (0.5,1.5) {\scriptsize $1$};
\node at (1.5,1.5) {\scriptsize $1$};
\end{tikzpicture} \
\begin{tikzpicture}[x=2.5mm, y=2.5mm]
\draw[black]
(0, 1) -- (0, 0) -- (1, 0) -- (2, 0) -- (2, 1) -- (2, 2) -- (1, 2) -- (0, 2) -- cycle;
\node at (0.5,0.5) {\scriptsize $1$};
\node at (1.5,0.5) {\scriptsize $2$};
\node at (0.5,1.5) {\scriptsize $2$};
\node at (1.5,1.5) {\scriptsize $1$};
\end{tikzpicture} \
\begin{tikzpicture}[x=2.5mm, y=2.5mm]
\draw[black]
(0, 1) -- (0, 0) -- (1, 0) -- (2, 0) -- (2, 1) -- (2, 2) -- (1, 2) -- (0, 2) -- cycle;
\node at (0.5,0.5) {\scriptsize $1$};
\node at (1.5,0.5) {\scriptsize $2$};
\node at (0.5,1.5) {\scriptsize $3$};
\node at (1.5,1.5) {\scriptsize $1$};
\end{tikzpicture} \
\begin{tikzpicture}[x=2.5mm, y=2.5mm]
\draw[black]
(0, 1) -- (0, 0) -- (1, 0) -- (2, 0) -- (2, 1) -- (2, 2) -- (1, 2) -- (0, 2) -- cycle;
\node at (0.5,0.5) {\scriptsize $1$};
\node at (1.5,0.5) {\scriptsize $3$};
\node at (0.5,1.5) {\scriptsize $1$};
\node at (1.5,1.5) {\scriptsize $2$};
\end{tikzpicture} \
\begin{tikzpicture}[x=2.5mm, y=2.5mm]
\draw[black]
(0, 1) -- (0, 0) -- (1, 0) -- (2, 0) -- (2, 1) -- (2, 2) -- (1, 2) -- (0, 2) -- cycle;
\node at (0.5,0.5) {\scriptsize $1$};
\node at (1.5,0.5) {\scriptsize $3$};
\node at (0.5,1.5) {\scriptsize $1$};
\node at (1.5,1.5) {\scriptsize $3$};
\end{tikzpicture} \
\begin{tikzpicture}[x=2.5mm, y=2.5mm]
\draw[black]
(0, 1) -- (0, 0) -- (1, 0) -- (2, 0) -- (2, 1) -- (2, 2) -- (1, 2) -- (0, 2) -- cycle;
\node at (0.5,0.5) {\scriptsize $1$};
\node at (1.5,0.5) {\scriptsize $3$};
\node at (0.5,1.5) {\scriptsize $2$};
\node at (1.5,1.5) {\scriptsize $2$};
\end{tikzpicture} \
\begin{tikzpicture}[x=2.5mm, y=2.5mm]
\draw[black]
(0, 1) -- (0, 0) -- (1, 0) -- (2, 0) -- (2, 1) -- (2, 2) -- (1, 2) -- (0, 2) -- cycle;
\node at (0.5,0.5) {\scriptsize $1$};
\node at (1.5,0.5) {\scriptsize $3$};
\node at (0.5,1.5) {\scriptsize $3$};
\node at (1.5,1.5) {\scriptsize $2$};
\end{tikzpicture} \
\begin{tikzpicture}[x=2.5mm, y=2.5mm]
\draw[black]
(0, 1) -- (0, 0) -- (1, 0) -- (2, 0) -- (2, 1) -- (2, 2) -- (1, 2) -- (0, 2) -- cycle;
\node at (0.5,0.5) {\scriptsize $1$};
\node at (1.5,0.5) {\scriptsize $3$};
\node at (0.5,1.5) {\scriptsize $3$};
\node at (1.5,1.5) {\scriptsize $3$};
\end{tikzpicture} \
\begin{tikzpicture}[x=2.5mm, y=2.5mm]
\draw[black]
(0, 1) -- (0, 0) -- (1, 0) -- (2, 0) -- (2, 1) -- (2, 2) -- (1, 2) -- (0, 2) -- cycle;
\node at (0.5,0.5) {\scriptsize $2$};
\node at (1.5,0.5) {\scriptsize $1$};
\node at (0.5,1.5) {\scriptsize $1$};
\node at (1.5,1.5) {\scriptsize $2$};
\end{tikzpicture} \
\begin{tikzpicture}[x=2.5mm, y=2.5mm]
\draw[black]
(0, 1) -- (0, 0) -- (1, 0) -- (2, 0) -- (2, 1) -- (2, 2) -- (1, 2) -- (0, 2) -- cycle;
\node at (0.5,0.5) {\scriptsize $2$};
\node at (1.5,0.5) {\scriptsize $1$};
\node at (0.5,1.5) {\scriptsize $1$};
\node at (1.5,1.5) {\scriptsize $3$};
\end{tikzpicture} \
\begin{tikzpicture}[x=2.5mm, y=2.5mm]
\draw[black]
(0, 1) -- (0, 0) -- (1, 0) -- (2, 0) -- (2, 1) -- (2, 2) -- (1, 2) -- (0, 2) -- cycle;
\node at (0.5,0.5) {\scriptsize $2$};
\node at (1.5,0.5) {\scriptsize $1$};
\node at (0.5,1.5) {\scriptsize $2$};
\node at (1.5,1.5) {\scriptsize $2$};
\end{tikzpicture}
}
$

$
\myvcenter{
\begin{tikzpicture}[x=2.5mm, y=2.5mm]
\draw[black]
(0, 1) -- (0, 0) -- (1, 0) -- (2, 0) -- (2, 1) -- (2, 2) -- (1, 2) -- (0, 2) -- cycle;
\node at (0.5,0.5) {\scriptsize $2$};
\node at (1.5,0.5) {\scriptsize $1$};
\node at (0.5,1.5) {\scriptsize $3$};
\node at (1.5,1.5) {\scriptsize $2$};
\end{tikzpicture} \
\begin{tikzpicture}[x=2.5mm, y=2.5mm]
\draw[black]
(0, 1) -- (0, 0) -- (1, 0) -- (2, 0) -- (2, 1) -- (2, 2) -- (1, 2) -- (0, 2) -- cycle;
\node at (0.5,0.5) {\scriptsize $2$};
\node at (1.5,0.5) {\scriptsize $1$};
\node at (0.5,1.5) {\scriptsize $3$};
\node at (1.5,1.5) {\scriptsize $3$};
\end{tikzpicture} \
\begin{tikzpicture}[x=2.5mm, y=2.5mm]
\draw[black]
(0, 1) -- (0, 0) -- (1, 0) -- (2, 0) -- (2, 1) -- (2, 2) -- (1, 2) -- (0, 2) -- cycle;
\node at (0.5,0.5) {\scriptsize $2$};
\node at (1.5,0.5) {\scriptsize $2$};
\node at (0.5,1.5) {\scriptsize $1$};
\node at (1.5,1.5) {\scriptsize $2$};
\end{tikzpicture} \
\begin{tikzpicture}[x=2.5mm, y=2.5mm]
\draw[black]
(0, 1) -- (0, 0) -- (1, 0) -- (2, 0) -- (2, 1) -- (2, 2) -- (1, 2) -- (0, 2) -- cycle;
\node at (0.5,0.5) {\scriptsize $2$};
\node at (1.5,0.5) {\scriptsize $2$};
\node at (0.5,1.5) {\scriptsize $1$};
\node at (1.5,1.5) {\scriptsize $3$};
\end{tikzpicture} \
\begin{tikzpicture}[x=2.5mm, y=2.5mm]
\draw[black]
(0, 1) -- (0, 0) -- (1, 0) -- (2, 0) -- (2, 1) -- (2, 2) -- (1, 2) -- (0, 2) -- cycle;
\node at (0.5,0.5) {\scriptsize $2$};
\node at (1.5,0.5) {\scriptsize $2$};
\node at (0.5,1.5) {\scriptsize $2$};
\node at (1.5,1.5) {\scriptsize $2$};
\end{tikzpicture} \
\begin{tikzpicture}[x=2.5mm, y=2.5mm]
\draw[black]
(0, 1) -- (0, 0) -- (1, 0) -- (2, 0) -- (2, 1) -- (2, 2) -- (1, 2) -- (0, 2) -- cycle;
\node at (0.5,0.5) {\scriptsize $2$};
\node at (1.5,0.5) {\scriptsize $2$};
\node at (0.5,1.5) {\scriptsize $3$};
\node at (1.5,1.5) {\scriptsize $2$};
\end{tikzpicture} \
\begin{tikzpicture}[x=2.5mm, y=2.5mm]
\draw[black]
(0, 1) -- (0, 0) -- (1, 0) -- (2, 0) -- (2, 1) -- (2, 2) -- (1, 2) -- (0, 2) -- cycle;
\node at (0.5,0.5) {\scriptsize $2$};
\node at (1.5,0.5) {\scriptsize $2$};
\node at (0.5,1.5) {\scriptsize $3$};
\node at (1.5,1.5) {\scriptsize $3$};
\end{tikzpicture} \
\begin{tikzpicture}[x=2.5mm, y=2.5mm]
\draw[black]
(0, 1) -- (0, 0) -- (1, 0) -- (2, 0) -- (2, 1) -- (2, 2) -- (1, 2) -- (0, 2) -- cycle;
\node at (0.5,0.5) {\scriptsize $3$};
\node at (1.5,0.5) {\scriptsize $1$};
\node at (0.5,1.5) {\scriptsize $2$};
\node at (1.5,1.5) {\scriptsize $1$};
\end{tikzpicture} \
\begin{tikzpicture}[x=2.5mm, y=2.5mm]
\draw[black]
(0, 1) -- (0, 0) -- (1, 0) -- (2, 0) -- (2, 1) -- (2, 2) -- (1, 2) -- (0, 2) -- cycle;
\node at (0.5,0.5) {\scriptsize $3$};
\node at (1.5,0.5) {\scriptsize $1$};
\node at (0.5,1.5) {\scriptsize $3$};
\node at (1.5,1.5) {\scriptsize $1$};
\end{tikzpicture} \
\begin{tikzpicture}[x=2.5mm, y=2.5mm]
\draw[black]
(0, 1) -- (0, 0) -- (1, 0) -- (2, 0) -- (2, 1) -- (2, 2) -- (1, 2) -- (0, 2) -- cycle;
\node at (0.5,0.5) {\scriptsize $3$};
\node at (1.5,0.5) {\scriptsize $2$};
\node at (0.5,1.5) {\scriptsize $2$};
\node at (1.5,1.5) {\scriptsize $1$};
\end{tikzpicture} \
\begin{tikzpicture}[x=2.5mm, y=2.5mm]
\draw[black]
(0, 1) -- (0, 0) -- (1, 0) -- (2, 0) -- (2, 1) -- (2, 2) -- (1, 2) -- (0, 2) -- cycle;
\node at (0.5,0.5) {\scriptsize $3$};
\node at (1.5,0.5) {\scriptsize $2$};
\node at (0.5,1.5) {\scriptsize $3$};
\node at (1.5,1.5) {\scriptsize $1$};
\end{tikzpicture} \
\begin{tikzpicture}[x=2.5mm, y=2.5mm]
\draw[black]
(0, 1) -- (0, 0) -- (1, 0) -- (2, 0) -- (2, 1) -- (2, 2) -- (1, 2) -- (0, 2) -- cycle;
\node at (0.5,0.5) {\scriptsize $3$};
\node at (1.5,0.5) {\scriptsize $3$};
\node at (0.5,1.5) {\scriptsize $2$};
\node at (1.5,1.5) {\scriptsize $2$};
\end{tikzpicture} \
\begin{tikzpicture}[x=2.5mm, y=2.5mm]
\draw[black]
(0, 1) -- (0, 0) -- (1, 0) -- (2, 0) -- (2, 1) -- (2, 2) -- (1, 2) -- (0, 2) -- cycle;
\node at (0.5,0.5) {\scriptsize $3$};
\node at (1.5,0.5) {\scriptsize $3$};
\node at (0.5,1.5) {\scriptsize $3$};
\node at (1.5,1.5) {\scriptsize $2$};
\end{tikzpicture} \
\begin{tikzpicture}[x=2.5mm, y=2.5mm]
\draw[black]
(0, 1) -- (0, 0) -- (1, 0) -- (2, 0) -- (2, 1) -- (2, 2) -- (1, 2) -- (0, 2) -- cycle;
\node at (0.5,0.5) {\scriptsize $3$};
\node at (1.5,0.5) {\scriptsize $3$};
\node at (0.5,1.5) {\scriptsize $3$};
\node at (1.5,1.5) {\scriptsize $3$};
\end{tikzpicture} \
}
$
\end{center}
(among the $81$ possible $2 \times 2$ patterns on three letters).
Equivalently, $\Ssurf$ is the set of elements of $\{1,2,3\}^{\bbZ^2}$
that do not any contain any of the following $11$ forbidden patterns.
\begin{center}
\myvcenter{%
\begin{tikzpicture}[x=2.5mm, y=2.5mm]
\draw[black]
(0, 1) -- (0, 0) -- (1, 0) -- (1, 1) -- (2, 1) -- (2, 2) -- (1, 2) -- (1, 1) -- cycle;
\node at (0.5,0.5) {\scriptsize $2$};
\node at (1.5,1.5) {\scriptsize $1$};
\end{tikzpicture} \
\begin{tikzpicture}[x=2.5mm, y=2.5mm]
\draw[black]
(0, 1) -- (0, 0) -- (1, 0) -- (1, 1) -- (1, 2) -- (0, 2) -- cycle;
\node at (0.5,0.5) {\scriptsize $3$};
\node at (0.5,1.5) {\scriptsize $1$};
\end{tikzpicture} \
\begin{tikzpicture}[x=2.5mm, y=2.5mm]
\draw[black]
(0, 1) -- (0, 0) -- (1, 0) -- (2, 0) -- (2, 1) -- (1, 1) -- cycle;
\node at (0.5,0.5) {\scriptsize $2$};
\node at (1.5,0.5) {\scriptsize $3$};
\end{tikzpicture} \
\begin{tikzpicture}[x=2.5mm, y=2.5mm]
\draw[black]
(0, 1) -- (0, 0) -- (1, 0) -- (2, 0) -- (2, 1) -- (2, 2) -- (1, 2) -- (1, 1) -- cycle;
\node at (0.5,0.5) {\scriptsize $1$};
\node at (1.5,0.5) {\scriptsize $1$};
\node at (1.5,1.5) {\scriptsize $2$};
\end{tikzpicture} \
\begin{tikzpicture}[x=2.5mm, y=2.5mm]
\draw[black]
(0, 1) -- (0, 0) -- (1, 0) -- (2, 0) -- (2, 1) -- (2, 2) -- (1, 2) -- (1, 1) -- cycle;
\node at (0.5,0.5) {\scriptsize $1$};
\node at (1.5,0.5) {\scriptsize $1$};
\node at (1.5,1.5) {\scriptsize $3$};
\end{tikzpicture} \
\begin{tikzpicture}[x=2.5mm, y=2.5mm]
\draw[black]
(0, 1) -- (0, 0) -- (1, 0) -- (2, 0) -- (2, 1) -- (2, 2) -- (1, 2) -- (1, 1) -- cycle;
\node at (0.5,0.5) {\scriptsize $1$};
\node at (1.5,0.5) {\scriptsize $2$};
\node at (1.5,1.5) {\scriptsize $2$};
\end{tikzpicture} \
\begin{tikzpicture}[x=2.5mm, y=2.5mm]
\draw[black]
(0, 1) -- (0, 0) -- (1, 0) -- (2, 0) -- (2, 1) -- (2, 2) -- (1, 2) -- (1, 1) -- cycle;
\node at (0.5,0.5) {\scriptsize $3$};
\node at (1.5,0.5) {\scriptsize $2$};
\node at (1.5,1.5) {\scriptsize $2$};
\end{tikzpicture} \
\begin{tikzpicture}[x=2.5mm, y=2.5mm]
\draw[black]
(0, 1) -- (0, 0) -- (1, 0) -- (2, 0) -- (2, 1) -- (2, 2) -- (1, 2) -- (1, 1) -- cycle;
\node at (0.5,0.5) {\scriptsize $3$};
\node at (1.5,0.5) {\scriptsize $1$};
\node at (1.5,1.5) {\scriptsize $3$};
\end{tikzpicture} \
\begin{tikzpicture}[x=2.5mm, y=2.5mm]
\draw[black]
(0, 1) -- (0, 0) -- (1, 0) -- (2, 0) -- (2, 1) -- (2, 2) -- (1, 2) -- (1, 1) -- cycle;
\node at (0.5,0.5) {\scriptsize $3$};
\node at (1.5,0.5) {\scriptsize $2$};
\node at (1.5,1.5) {\scriptsize $3$};
\end{tikzpicture} \
\begin{tikzpicture}[x=2.5mm, y=2.5mm]
\draw[black]
(0, 1) -- (0, 0) -- (1, 0) -- (2, 0) -- (2, 1) -- (2, 2) -- (1, 2) -- (1, 1) -- cycle;
\node at (0.5,0.5) {\scriptsize $1$};
\node at (1.5,0.5) {\scriptsize $2$};
\node at (1.5,1.5) {\scriptsize $3$};
\end{tikzpicture} \
\begin{tikzpicture}[x=2.5mm, y=2.5mm]
\draw[black]
(0, 1) -- (0, 0) -- (1, 0) -- (2, 0) -- (2, 1) -- (2, 2) -- (1, 2) -- (1, 1) -- cycle;
\node at (0.5,0.5) {\scriptsize $3$};
\node at (1.5,0.5) {\scriptsize $1$};
\node at (1.5,1.5) {\scriptsize $2$};
\end{tikzpicture}
}
\end{center}
It is proved in \cite{Jam04} that the elements of $\Ssurf$
correspond to codings of \emph{stepped surfaces} in $\bbR^3$,
as they are defined in \cite{ABFJ07}.
Some of the most interesting known examples of combinatorial substitutions
correspond to \emph{dual substitutions} (introduced in \cite{AI01}, see also \cite{ABFJ07}),
which are substitutions acting on the unit faces of stepped surfaces.
We are not going to describe dual substitutions any further,
but we will only consider the two-dimensional combinatorial substitutions that they give rise to,
forgetting about where they come from.

The theory developed in \cite{AI01} enables us to prove that
every combinatorial substitution that comes from
a dual substitution is consistent and non-overlapping,
but the proof is not combinatorial
and relies on specific arguments about discrete surfaces and dual substitutions.
The authors of \cite{ABS04} have asked for a ``purely combinatorial'' proof
in the particular case of the substitution of Example~\ref{exam:mini}:
``Unfortunately, although this theorem appears to be a purely combinatorial result,
we do not know any combinatorial proof of it, and we would be very interested in such a proof.''
Such a proof has also been asked for in \cite{PF02}.
We will now describe how the results of Section~\ref{sect:dec} can be used
to answer the questions raised above.

\paragraph{Consistency in $\Ssurf$}
The consistency of a $\Psurf$-domino-complete substitution on $\patt{\Ssurf}$,
can be checked simply by checking its consistency on the $28$ patterns in $\Psurf$,
thanks to Theorem~\ref{theo:decconsmieux}.

\paragraph{Overlapping in $\Ssurf$}
Checking the overlapping property on $\patt(\Ssurf)$
for $\Psurf$-domino-complete substitutions is not as easy as checking consistency,
because we haven't proved an ``enhanced'' version of Theorem~\ref{theo:decnono}
as we did with Theorem~\ref{theo:decconsmieux} for Theorem~\ref{theo:deccons}.

\begin{exam}
\label{exam:mini}
Let $\sigma$ be the two-dimensional substitution on alphabet $\{1,2,3\}$ defined by
the base rule
$1 \mapsto
\myvcenter{%
\begin{tikzpicture}[x=2.5mm, y=2.5mm]
\draw[black]
(0, 1) -- (0, 0) -- (1, 0) -- (1, 1) -- (1, 2) -- (0, 2) -- cycle;
\node at (0.5,0.5) {\scriptsize $2$};
\node at (0.5,1.5) {\scriptsize $1$};
\end{tikzpicture}}$,
$2 \mapsto
\myvcenter{%
\begin{tikzpicture}[x=2.5mm, y=2.5mm]
\draw[black]
(0, 0) -- (1, 0) -- (1, 1) -- (0, 1) -- cycle;
\node at (0.5,0.5) {\scriptsize $3$};
\end{tikzpicture}}$,
$3 \mapsto
\myvcenter{%
\begin{tikzpicture}[x=2.5mm, y=2.5mm]
\draw[black]
(0, 0) -- (1, 0) -- (1, 1) -- (0, 1) -- cycle;
\node at (0.5,0.5) {\scriptsize $1$};
\end{tikzpicture}}$
and the concatenation rules
\begin{center}
$
\begin{array}{rclcrclcrclcrcl}
\myvcenter{\begin{tikzpicture}[x=2.5mm, y=2.5mm]
\draw[black]
(0, 1) -- (0, 0) -- (1, 0) -- (1, 1) -- cycle;
\node at (0.5,0.5) {\scriptsize $1$};
\definecolor{cellcolor}{rgb}{0.65,0.65,0.65}
\fill[fill=cellcolor, draw=black]
(1, 1) -- (1, 0) -- (2, 0) -- (2, 1) -- cycle;
\node at (1.5,0.5) {\scriptsize $1$};
\end{tikzpicture}}
& \mapsto &
\myvcenter{\begin{tikzpicture}[x=2.5mm, y=2.5mm]
\draw[black]
(0, 0) -- (0, -1) -- (0, -2) -- (1, -2) -- (1, -1) -- (1, 0) -- cycle;
\node at (0.5,-1.5) {\scriptsize $2$};
\node at (0.5,-0.5) {\scriptsize $1$};
\definecolor{cellcolor}{rgb}{0.65,0.65,0.65}
\fill[fill=cellcolor, draw=black]
(0, 1) -- (0, 0) -- (1, 0) -- (1, 1) -- (1, 2) -- (0, 2) -- cycle;
\node at (0.5,0.5) {\scriptsize $2$};
\node at (0.5,1.5) {\scriptsize $1$};
\end{tikzpicture}}
& \quad &
\myvcenter{\begin{tikzpicture}[x=2.5mm, y=2.5mm]
\draw[black]
(0, 1) -- (0, 0) -- (1, 0) -- (1, 1) -- cycle;
\node at (0.5,0.5) {\scriptsize $1$};
\definecolor{cellcolor}{rgb}{0.65,0.65,0.65}
\fill[fill=cellcolor, draw=black]
(1, 1) -- (1, 0) -- (2, 0) -- (2, 1) -- cycle;
\node at (1.5,0.5) {\scriptsize $2$};
\end{tikzpicture}}
& \mapsto &
\myvcenter{\begin{tikzpicture}[x=2.5mm, y=2.5mm]
\draw[black]
(0, 1) -- (0, 0) -- (1, 0) -- (1, 1) -- (1, 2) -- (0, 2) -- cycle;
\node at (0.5,0.5) {\scriptsize $2$};
\node at (0.5,1.5) {\scriptsize $1$};
\definecolor{cellcolor}{rgb}{0.65,0.65,0.65}
\fill[fill=cellcolor, draw=black]
(0, 3) -- (0, 2) -- (1, 2) -- (1, 3) -- cycle;
\node at (0.5,2.5) {\scriptsize $3$};
\end{tikzpicture}}
& \quad &
\myvcenter{\begin{tikzpicture}[x=2.5mm, y=2.5mm]
\draw[black]
(0, 1) -- (0, 0) -- (1, 0) -- (1, 1) -- cycle;
\node at (0.5,0.5) {\scriptsize $1$};
\definecolor{cellcolor}{rgb}{0.65,0.65,0.65}
\fill[fill=cellcolor, draw=black]
(1, 1) -- (1, 0) -- (2, 0) -- (2, 1) -- cycle;
\node at (1.5,0.5) {\scriptsize $3$};
\end{tikzpicture}}
& \mapsto &
\myvcenter{\begin{tikzpicture}[x=2.5mm, y=2.5mm]
\draw[black]
(0, 0) -- (0, -1) -- (0, -2) -- (1, -2) -- (1, -1) -- (1, 0) -- cycle;
\node at (0.5,-1.5) {\scriptsize $2$};
\node at (0.5,-0.5) {\scriptsize $1$};
\definecolor{cellcolor}{rgb}{0.65,0.65,0.65}
\fill[fill=cellcolor, draw=black]
(0, 1) -- (0, 0) -- (1, 0) -- (1, 1) -- cycle;
\node at (0.5,0.5) {\scriptsize $1$};
\end{tikzpicture}}
& \quad &
\myvcenter{\begin{tikzpicture}[x=2.5mm, y=2.5mm]
\draw[black]
(0, 1) -- (0, 0) -- (1, 0) -- (1, 1) -- cycle;
\node at (0.5,0.5) {\scriptsize $2$};
\definecolor{cellcolor}{rgb}{0.65,0.65,0.65}
\fill[fill=cellcolor, draw=black]
(1, 1) -- (1, 0) -- (2, 0) -- (2, 1) -- cycle;
\node at (1.5,0.5) {\scriptsize $1$};
\end{tikzpicture}}
& \mapsto &
\myvcenter{\begin{tikzpicture}[x=2.5mm, y=2.5mm]
\draw[black]
(0, 0) -- (0, -1) -- (1, -1) -- (1, 0) -- cycle;
\node at (0.5,-0.5) {\scriptsize $3$};
\definecolor{cellcolor}{rgb}{0.65,0.65,0.65}
\fill[fill=cellcolor, draw=black]
(0, 1) -- (0, 0) -- (1, 0) -- (1, 1) -- (1, 2) -- (0, 2) -- cycle;
\node at (0.5,0.5) {\scriptsize $2$};
\node at (0.5,1.5) {\scriptsize $1$};
\end{tikzpicture}}
\\
\myvcenter{\begin{tikzpicture}[x=2.5mm, y=2.5mm]
\draw[black]
(0, 1) -- (0, 0) -- (1, 0) -- (1, 1) -- cycle;
\node at (0.5,0.5) {\scriptsize $2$};
\definecolor{cellcolor}{rgb}{0.65,0.65,0.65}
\fill[fill=cellcolor, draw=black]
(1, 1) -- (1, 0) -- (2, 0) -- (2, 1) -- cycle;
\node at (1.5,0.5) {\scriptsize $2$};
\end{tikzpicture}}
& \mapsto &
\myvcenter{\begin{tikzpicture}[x=2.5mm, y=2.5mm]
\draw[black]
(0, 0) -- (0, -1) -- (1, -1) -- (1, 0) -- cycle;
\node at (0.5,-0.5) {\scriptsize $3$};
\definecolor{cellcolor}{rgb}{0.65,0.65,0.65}
\fill[fill=cellcolor, draw=black]
(0, 1) -- (0, 0) -- (1, 0) -- (1, 1) -- cycle;
\node at (0.5,0.5) {\scriptsize $3$};
\end{tikzpicture}}
& \quad &
\myvcenter{\begin{tikzpicture}[x=2.5mm, y=2.5mm]
\draw[black]
(0, 1) -- (0, 0) -- (1, 0) -- (1, 1) -- cycle;
\node at (0.5,0.5) {\scriptsize $3$};
\definecolor{cellcolor}{rgb}{0.65,0.65,0.65}
\fill[fill=cellcolor, draw=black]
(1, 1) -- (1, 0) -- (2, 0) -- (2, 1) -- cycle;
\node at (1.5,0.5) {\scriptsize $1$};
\end{tikzpicture}}
& \mapsto &
\myvcenter{\begin{tikzpicture}[x=2.5mm, y=2.5mm]
\draw[black]
(0, 1) -- (0, 0) -- (1, 0) -- (1, 1) -- cycle;
\node at (0.5,0.5) {\scriptsize $1$};
\definecolor{cellcolor}{rgb}{0.65,0.65,0.65}
\fill[fill=cellcolor, draw=black]
(0, 2) -- (0, 1) -- (1, 1) -- (1, 2) -- (1, 3) -- (0, 3) -- cycle;
\node at (0.5,1.5) {\scriptsize $2$};
\node at (0.5,2.5) {\scriptsize $1$};
\end{tikzpicture}}
& \quad &
\myvcenter{\begin{tikzpicture}[x=2.5mm, y=2.5mm]
\draw[black]
(0, 1) -- (0, 0) -- (1, 0) -- (1, 1) -- cycle;
\node at (0.5,0.5) {\scriptsize $3$};
\definecolor{cellcolor}{rgb}{0.65,0.65,0.65}
\fill[fill=cellcolor, draw=black]
(1, 1) -- (1, 0) -- (2, 0) -- (2, 1) -- cycle;
\node at (1.5,0.5) {\scriptsize $2$};
\end{tikzpicture}}
& \mapsto &
\myvcenter{\begin{tikzpicture}[x=2.5mm, y=2.5mm]
\draw[black]
(0, 1) -- (0, 0) -- (1, 0) -- (1, 1) -- cycle;
\node at (0.5,0.5) {\scriptsize $1$};
\definecolor{cellcolor}{rgb}{0.65,0.65,0.65}
\fill[fill=cellcolor, draw=black]
(0, 2) -- (0, 1) -- (1, 1) -- (1, 2) -- cycle;
\node at (0.5,1.5) {\scriptsize $3$};
\end{tikzpicture}}
& \quad &
\myvcenter{\begin{tikzpicture}[x=2.5mm, y=2.5mm]
\draw[black]
(0, 1) -- (0, 0) -- (1, 0) -- (1, 1) -- cycle;
\node at (0.5,0.5) {\scriptsize $3$};
\definecolor{cellcolor}{rgb}{0.65,0.65,0.65}
\fill[fill=cellcolor, draw=black]
(1, 1) -- (1, 0) -- (2, 0) -- (2, 1) -- cycle;
\node at (1.5,0.5) {\scriptsize $3$};
\end{tikzpicture}}
& \mapsto &
\myvcenter{\begin{tikzpicture}[x=2.5mm, y=2.5mm]
\draw[black]
(0, 1) -- (0, 0) -- (1, 0) -- (1, 1) -- cycle;
\node at (0.5,0.5) {\scriptsize $1$};
\definecolor{cellcolor}{rgb}{0.65,0.65,0.65}
\fill[fill=cellcolor, draw=black]
(0, 2) -- (0, 1) -- (1, 1) -- (1, 2) -- cycle;
\node at (0.5,1.5) {\scriptsize $1$};
\end{tikzpicture}}
\\
\myvcenter{\begin{tikzpicture}[x=2.5mm, y=2.5mm]
\draw[black]
(0, 1) -- (0, 0) -- (1, 0) -- (1, 1) -- cycle;
\node at (0.5,0.5) {\scriptsize $1$};
\definecolor{cellcolor}{rgb}{0.65,0.65,0.65}
\fill[fill=cellcolor, draw=black]
(0, 2) -- (0, 1) -- (1, 1) -- (1, 2) -- cycle;
\node at (0.5,1.5) {\scriptsize $1$};
\end{tikzpicture}}
& \mapsto &
\myvcenter{\begin{tikzpicture}[x=2.5mm, y=2.5mm]
\draw[black]
(1, 2) -- (1, 1) -- (2, 1) -- (2, 2) -- (2, 3) -- (1, 3) -- cycle;
\node at (1.5,1.5) {\scriptsize $2$};
\node at (1.5,2.5) {\scriptsize $1$};
\definecolor{cellcolor}{rgb}{0.65,0.65,0.65}
\fill[fill=cellcolor, draw=black]
(0, 1) -- (0, 0) -- (1, 0) -- (1, 1) -- (1, 2) -- (0, 2) -- cycle;
\node at (0.5,0.5) {\scriptsize $2$};
\node at (0.5,1.5) {\scriptsize $1$};
\end{tikzpicture}}
& \quad &
\myvcenter{\begin{tikzpicture}[x=2.5mm, y=2.5mm]
\draw[black]
(0, 1) -- (0, 0) -- (1, 0) -- (1, 1) -- cycle;
\node at (0.5,0.5) {\scriptsize $1$};
\definecolor{cellcolor}{rgb}{0.65,0.65,0.65}
\fill[fill=cellcolor, draw=black]
(0, 2) -- (0, 1) -- (1, 1) -- (1, 2) -- cycle;
\node at (0.5,1.5) {\scriptsize $2$};
\end{tikzpicture}}
& \mapsto &
\myvcenter{\begin{tikzpicture}[x=2.5mm, y=2.5mm]
\draw[black]
(1, 1) -- (1, 0) -- (2, 0) -- (2, 1) -- (2, 2) -- (1, 2) -- cycle;
\node at (1.5,0.5) {\scriptsize $2$};
\node at (1.5,1.5) {\scriptsize $1$};
\definecolor{cellcolor}{rgb}{0.65,0.65,0.65}
\fill[fill=cellcolor, draw=black]
(0, 1) -- (0, 0) -- (1, 0) -- (1, 1) -- cycle;
\node at (0.5,0.5) {\scriptsize $3$};
\end{tikzpicture}}
& \quad &
\myvcenter{\begin{tikzpicture}[x=2.5mm, y=2.5mm]
\draw[black]
(0, 1) -- (0, 0) -- (1, 0) -- (1, 1) -- cycle;
\node at (0.5,0.5) {\scriptsize $1$};
\definecolor{cellcolor}{rgb}{0.65,0.65,0.65}
\fill[fill=cellcolor, draw=black]
(0, 2) -- (0, 1) -- (1, 1) -- (1, 2) -- cycle;
\node at (0.5,1.5) {\scriptsize $3$};
\end{tikzpicture}}
& \mapsto &
\myvcenter{\begin{tikzpicture}[x=2.5mm, y=2.5mm]
\draw[black]
(0, 1) -- (0, 0) -- (1, 0) -- (1, 1) -- (1, 2) -- (0, 2) -- cycle;
\node at (0.5,0.5) {\scriptsize $2$};
\node at (0.5,1.5) {\scriptsize $1$};
\definecolor{cellcolor}{rgb}{0.65,0.65,0.65}
\fill[fill=cellcolor, draw=black]
(-1, 1) -- (-1, 0) -- (0, 0) -- (0, 1) -- cycle;
\node at (-0.5,0.5) {\scriptsize $1$};
\end{tikzpicture}}
& \quad &
\myvcenter{\begin{tikzpicture}[x=2.5mm, y=2.5mm]
\draw[black]
(0, 1) -- (0, 0) -- (1, 0) -- (1, 1) -- cycle;
\node at (0.5,0.5) {\scriptsize $2$};
\definecolor{cellcolor}{rgb}{0.65,0.65,0.65}
\fill[fill=cellcolor, draw=black]
(0, 2) -- (0, 1) -- (1, 1) -- (1, 2) -- cycle;
\node at (0.5,1.5) {\scriptsize $1$};
\end{tikzpicture}}
& \mapsto &
\myvcenter{\begin{tikzpicture}[x=2.5mm, y=2.5mm]
\draw[black]
(1, 2) -- (1, 1) -- (2, 1) -- (2, 2) -- cycle;
\node at (1.5,1.5) {\scriptsize $3$};
\definecolor{cellcolor}{rgb}{0.65,0.65,0.65}
\fill[fill=cellcolor, draw=black]
(0, 1) -- (0, 0) -- (1, 0) -- (1, 1) -- (1, 2) -- (0, 2) -- cycle;
\node at (0.5,0.5) {\scriptsize $2$};
\node at (0.5,1.5) {\scriptsize $1$};
\end{tikzpicture}}
\\
\myvcenter{\begin{tikzpicture}[x=2.5mm, y=2.5mm]
\draw[black]
(0, 1) -- (0, 0) -- (1, 0) -- (1, 1) -- cycle;
\node at (0.5,0.5) {\scriptsize $2$};
\definecolor{cellcolor}{rgb}{0.65,0.65,0.65}
\fill[fill=cellcolor, draw=black]
(0, 2) -- (0, 1) -- (1, 1) -- (1, 2) -- cycle;
\node at (0.5,1.5) {\scriptsize $2$};
\end{tikzpicture}}
& \mapsto &
\myvcenter{\begin{tikzpicture}[x=2.5mm, y=2.5mm]
\draw[black]
(1, 1) -- (1, 0) -- (2, 0) -- (2, 1) -- cycle;
\node at (1.5,0.5) {\scriptsize $3$};
\definecolor{cellcolor}{rgb}{0.65,0.65,0.65}
\fill[fill=cellcolor, draw=black]
(0, 1) -- (0, 0) -- (1, 0) -- (1, 1) -- cycle;
\node at (0.5,0.5) {\scriptsize $3$};
\end{tikzpicture}}
& \quad &
\myvcenter{\begin{tikzpicture}[x=2.5mm, y=2.5mm]
\draw[black]
(0, 1) -- (0, 0) -- (1, 0) -- (1, 1) -- cycle;
\node at (0.5,0.5) {\scriptsize $2$};
\definecolor{cellcolor}{rgb}{0.65,0.65,0.65}
\fill[fill=cellcolor, draw=black]
(0, 2) -- (0, 1) -- (1, 1) -- (1, 2) -- cycle;
\node at (0.5,1.5) {\scriptsize $3$};
\end{tikzpicture}}
& \mapsto &
\myvcenter{\begin{tikzpicture}[x=2.5mm, y=2.5mm]
\draw[black]
(0, 1) -- (0, 0) -- (1, 0) -- (1, 1) -- cycle;
\node at (0.5,0.5) {\scriptsize $3$};
\definecolor{cellcolor}{rgb}{0.65,0.65,0.65}
\fill[fill=cellcolor, draw=black]
(-1, 1) -- (-1, 0) -- (0, 0) -- (0, 1) -- cycle;
\node at (-0.5,0.5) {\scriptsize $1$};
\end{tikzpicture}}
& \quad &
\myvcenter{\begin{tikzpicture}[x=2.5mm, y=2.5mm]
\draw[black]
(0, 1) -- (0, 0) -- (1, 0) -- (1, 1) -- cycle;
\node at (0.5,0.5) {\scriptsize $3$};
\definecolor{cellcolor}{rgb}{0.65,0.65,0.65}
\fill[fill=cellcolor, draw=black]
(0, 2) -- (0, 1) -- (1, 1) -- (1, 2) -- cycle;
\node at (0.5,1.5) {\scriptsize $2$};
\end{tikzpicture}}
& \mapsto &
\myvcenter{\begin{tikzpicture}[x=2.5mm, y=2.5mm]
\draw[black]
(0, 1) -- (0, 0) -- (1, 0) -- (1, 1) -- cycle;
\node at (0.5,0.5) {\scriptsize $1$};
\definecolor{cellcolor}{rgb}{0.65,0.65,0.65}
\fill[fill=cellcolor, draw=black]
(-1, 0) -- (-1, -1) -- (0, -1) -- (0, 0) -- cycle;
\node at (-0.5,-0.5) {\scriptsize $3$};
\end{tikzpicture}}
& \quad &
\myvcenter{\begin{tikzpicture}[x=2.5mm, y=2.5mm]
\draw[black]
(0, 1) -- (0, 0) -- (1, 0) -- (1, 1) -- cycle;
\node at (0.5,0.5) {\scriptsize $3$};
\definecolor{cellcolor}{rgb}{0.65,0.65,0.65}
\fill[fill=cellcolor, draw=black]
(0, 2) -- (0, 1) -- (1, 1) -- (1, 2) -- cycle;
\node at (0.5,1.5) {\scriptsize $3$};
\end{tikzpicture}}
& \mapsto &
\myvcenter{\begin{tikzpicture}[x=2.5mm, y=2.5mm]
\draw[black]
(0, 1) -- (0, 0) -- (1, 0) -- (1, 1) -- cycle;
\node at (0.5,0.5) {\scriptsize $1$};
\definecolor{cellcolor}{rgb}{0.65,0.65,0.65}
\fill[fill=cellcolor, draw=black]
(-1, 0) -- (-1, -1) -- (0, -1) -- (0, 0) -- cycle;
\node at (-0.5,-0.5) {\scriptsize $1$};
\end{tikzpicture}}.
\end{array}
$
\end{center}
This substitution is the main object of study of \cite{ABS04} and it also appears in \cite{Fer07}.
It is $\Psurf$-domino-complete,
and it is consistent and non-overlapping on $\patt(\Ssurf)$.
\end{exam}

\begin{prop}
The substitution of Example~\ref{exam:mini} is consistent and non-overlapping on $\patt(\Ssurf)$.
\end{prop}

\begin{proof}
Consistency can be checked algorithmically, thanks to Theorem~\ref{theo:decconsmieux}.
To prove that it is non-overlapping,
let $c = [(0,0),t]$ and $c'=[(x,y),t']$ be two distinct cells belonging to a pattern $P \in \patt(\Ssurf)$.
There exists a $\mcCs$-path from $c$ to $c'$ that consists of a horizontal segment followed by a vertical segment,
because $\sigma$ is $\Psurf$-domino-complete and $P$ is extendible to an element of $\Ssurf$.
Along the horizontal segment, all image vectors are of the form $(0,i)$,
because the right-hand sides of the rules of $\sigma$ for horizontal dominoes
have the image patterns aligned vertically.
Along the vertical segment, all image vectors are of the form $(-1,i)$ when moving upward,
because the images of the vertical dominoes are aligned that way.
It follows that $\ws(\gamma) = (-y,i)$ with $i \in \bbZ$.
Since the image patterns have width $1$,
it is clear that there is no overlap between the images of $c$ and $c'$ if $y \neq 0$.
If $y=0$ then we are considering a horizontal path,
and it is clear that such path induces no overlap either because the images are stacked on top of each other.
\end{proof}

\section{Conclusion and open problems}
In this article we focused on the consistency and the overlapping of substitutions,
because they are the first properties that one should look at,
in order to make sure that the substitution is ``well defined''.
There are many other interesting properties whose decidability status
has not been investigated yet. Some of them are:

\begin{itemize}
\itemsep=0pt \parskip=0pt
\item
The overlapping property in the consistent and ``weakened'' domino-complete case,
as we have done for consistency in Theorem~\ref{theo:decconsmieux}.
\item
Iterability:
given a substitution $\sigma$ and a pattern $P$,
can $\sigma$ be iterated on the successive images of $P$ by $\sigma$?
(That is, are the successive images of $P$ by $\sigma$ all $\mcCs$-covered?)
\item
Consistency in the primitive case.
A substitution is \emph{primitive} if for every cell of any type,
there exists $n \in \bbN$ such that the pattern $\sigma^n(\{c\})$
contain cells of \emph{every} type of $\mcA$.
(This presupposes that the substitution can be iterated.)
Is it decidable if a primitive substitution is consistent?
\item
(Simple) connectedness:
given a substitution $\sigma$ and a pattern $P$,
are the iterates of $\sigma$ on $P$ (simply) connected?
\item
The ``growing squares'' property:
given a substitution $\sigma$ and a pattern $P$,
do the iterates of $\sigma$ on $P$ contain arbitrarily large squares?
\end{itemize}

\section*{Acknowledgements}
Many thanks to Thomas Fernique for helpful discussions.
Research supported by the Academy of Finland Grant 131558.

\bibliographystyle{amsalpha}
\bibliography{biblio}
\end{document}